\documentclass[utf8]{frontiersSCNS} 

\usepackage{url,hyperref,lineno,microtype,subcaption}
\usepackage[onehalfspacing]{setspace}
\usepackage[normalem]{ulem}

\renewcommand{\vec}[1]{\boldsymbol{#1}} 

\def\keyFont{\fontsize{8}{11}\helveticabold }
\def\firstAuthorLast{Verscharen {et~al.}} 
\def\Authors{Daniel Verscharen\,$^{1,*}$, Chandran, B.~D.~G.\,$^{2,3}$, Boella, E.\,$^{4,5}$, Halekas, J.\,$^{6}$, Innocenti, M.~E.\,$^{7}$, Jagarlamudi, V.~K.\,$^{8,9}$, Micera, A.\,$^{10}$, Pierrard, V.\,$^{10,11}$, {\v{S}}tver\'ak, {\v{S}}.\,$^{12,13}$, Vasko, I.~Y.\,$^{14,15}$, Velli, M.\,$^{16}$ and Whittlesey, P.~L.\,$^{14}$ }
\def\Address{$^{1}$Mullard Space Science Laboratory, University College London, Dorking, UK \\
$^{2}$Space Science Center, University of New Hampshire, Durham NH, USA\\
$^{3}$Department of Physics and Astronomy, University of New Hampshire, Durham NH, USA\\
$^{4}$Physics Department, Lancaster University, Lancaster, UK\\
$^{5}$Cockroft Institute, Daresbury Laboratory, Warrington, UK\\
$^{6}$Department of Physics and Astronomy, University of Iowa, Iowa City IA, USA\\
$^{7}$Institut f\"ur Theoretische Physik, Ruhr-Universit\"at Bochum, Bochum, Germany\\
$^{8}$Applied Physics Laboratory, Johns Hopkins University, Laurel MD, USA\\
$^{9}$National Institute for Astrophysics -- Institute for Space Astrophysics and Planetology, Rome, Italy\\
$^{10}$Solar-Terrestrial Centre of Excellence, Royal Observatory of Belgium, Brussels, Belgium\\
$^{11}$Center for Space Radiations (CSR) and Georges Lema\^{i}tre Centre for Earth and Climate Research (TECLIM), Earth and Life Institute (ELI), Universit\'e Catholique de Louvain (UCLouvain), Louvain-La-Neuve, Belgium\\
$^{12}$Institute of Atmospheric Physics of the Czech Academy of Sciences, Prague, Czech Republic\\
$^{13}$Astronomical Institute of the Czech Academy of Sciences, Prague, Czech Republic\\
$^{14}$Space Sciences Laboratory, University of California at Berkeley, Berkeley CA, USA\\
$^{15}$Space Research Institute, Russian Academy of Sciences, Moscow, Russia\\
$^{16}$Department of Earth, Planetary, and Space Sciences, University of California Los Angeles, Los Angeles CA, USA}
\def\corrAuthor{Daniel Verscharen}

\def\corrEmail{d.verscharen@ucl.ac.uk}

\begin{document}
\onecolumn
\firstpage{1}

\title[Electron-driven instabilities]{Electron-driven instabilities in the solar wind} 

\author[\firstAuthorLast ]{\Authors} 
\address{} 
\correspondance{} 

\extraAuth{}

\maketitle

\begin{abstract}

\section{}
The electrons are an essential particle species in the solar wind. They often exhibit non-equilibrium features in their velocity distribution function. These include temperature anisotropies,  tails (kurtosis), and reflectional asymmetries (skewness), which contribute a significant heat flux to the solar wind. If these non-equilibrium features are sufficiently strong, they drive kinetic micro-instabilities. We develop a semi-graphical framework based on the equations of quasi-linear theory to describe electron-driven instabilities in the solar wind. We apply our framework to resonant instabilities driven by temperature anisotropies. These include the electron whistler anisotropy instability 
and the propagating electron firehose instability. We then describe resonant instabilities driven by reflectional asymmetries in the electron distribution function. These include the electron/ion-acoustic, kinetic Alfv\'en heat-flux, Langmuir, electron-beam, electron/ion-cyclotron, electron/electron-acoustic, whistler heat-flux, oblique fast-magnetosonic/whistler, lower-hybrid fan, and electron-deficit whistler instability. We briefly comment on  non-resonant instabilities driven by electron temperature anisotropies such as the mirror-mode and the non-propagating firehose instability. We conclude our review with a list of open research topics in the field of electron-driven instabilities in the solar wind.

\tiny
 \keyFont{ \section{Keywords:} solar wind, plasma, instabilities, electrons, temperature anisotropy, heat flux, quasi-linear theory} 
\end{abstract}

\section{Introduction}

The solar wind is a fully ionised and quasi-neutral plasma flow \citep[for a recent review about the solar wind, see][]{verscharen19a}. Plasma flows with these properties consist of free negatively charged electrons and free positively charged ions. The majority of the ions in the solar wind are protons with an addition of 2-5\% of $\alpha$-particles and a minority contribution of heavier ions. Quasi-neutrality requires that electrons and ions are spatially distributed so that the total charge density of the plasma is approximately zero on scales much greater than the Debye length. In order to fulfill quasi-neutrality, electrons must be, on average, the particle species with the greatest number density in fully ionised and quasi-neutral plasmas like the solar wind. 

The mass of an electron is by a factor of 1836 times smaller than the mass of a proton. Therefore, the direct contributions of electrons to the solar-wind mass, momentum, angular-momentum, and kinetic-energy fluxes are negligible compared to the proton contributions. However, electrons contribute significantly to the overall momentum balance of the solar wind through their thermal pressure gradient \citep{parker58,landi03} and to the overall energy balance of the solar wind through their heat flux \citep{hollweg74,scime99,pagel05,scime94,bale13,borovsky14,cranmer21}. This is true both for fast solar-wind streams, whose sources are open coronal field regions such as polar coronal holes, as well as for the wind originating from the more complex coronal regions associated with helmet streamers and pseudo-streamers. In the simplest models of coronal acceleration, the fluid electron pressure gradient reflects the effects of the interplanetary electric field set up by the much greater scale height of electrons compared to protons of similar temperatures \citep{parker_kinetic_2010}. The subsequent (Jeans-theorem) evolution of the wind, taking into account charge conservation (the outflow must be globally neutral) as well as local charge neutrality, together with magnetic-moment conservation for particles of each species, leads to distribution functions in the supersonic wind that are strongly out of equilibrium. These distributions become  unstable to plasma and electromagnetic field oscillations that most likely play a major role in shaping the observed distributions as we discuss in this article. 

Electron-kinetic processes such as resonant damping and instabilities modify the overall energy budget of the electromagnetic plasma fluctuations, which has an impact on the overall evolution of the solar wind \citep{gary75a,feldman76a,ramani78,gary99,alexandrova09,schekochihin09,stverak15}. Estimates of the empirical proton-to-total heating ratio based on observed temperature profiles in the inner heliosphere suggest that a significant fraction ($\sim 40\%$) of the turbulent energy is dissipated by electrons \citep{cranmer09}. Therefore, electrons and electron-driven processes are considered essential for our understanding of the global evolution of the solar wind \citep[and for other astrophysical plasmas, see][]{verscharen21,verscharen21b}.

In-situ solar-wind measurements show that the electrons, like the ions, often exhibit deviations from thermodynamic equilibrium \citep{feldman75,rosenbauer77,pilip87,maksimovic97}. These deviations become apparent in the electrons' velocity distribution function $f_{\mathrm e}$ that often differs from the Maxwellian equilibrium distribution. We define $f_{\mathrm e}$ so that $f_{\mathrm e}(\vec x,\vec v,t)\,\mathrm d^3x\,\mathrm d^3v$ describes the total number of electrons in the phase-space volume $\mathrm d^3x\,\mathrm d^3v$ centred around the coordinates $(\vec x,\vec v)$ at time~$t$. If binary Coulomb collisions between the plasma particles were the dominant process that determined~$f_{\mathrm e}$, the observed deviations from the Maxwellian equilibrium would not persist, at least not over long timescales when compared to the Coulomb collision time. We, therefore, refer to the solar wind often as a \emph{collisionless plasma} \citep{marsch06}. Given the steep energy dependence of the Coulomb-collision cross section, this applies especially to the suprathermal electrons; however, for the thermal electrons, collisions remain  important \citep{scudder_theory_1979, landi_competition_2012}.  

Temperature anisotropy is a typical non-thermal feature associated with $f_{\mathrm e}$ in the solar wind \citep{phillips89,salem03,stverak08}. Temperature anisotropy is characterised by different temperatures in the directions perpendicular and parallel to the local magnetic field. In this context, we understand temperature as the \emph{kinetic temperature} based on the diagonal elements of the electron pressure tensor (i.e., the second velocity moment of $f_{\mathrm e}$). We define the temperature of a plasma species $j$ in the direction perpendicular to the magnetic field as $T_{\perp j}$ and its temperature in the direction parallel to the magnetic field as $T_{\parallel j}$. 

\begin{figure}[ht]
\begin{center}
\includegraphics[width=0.8\textwidth]{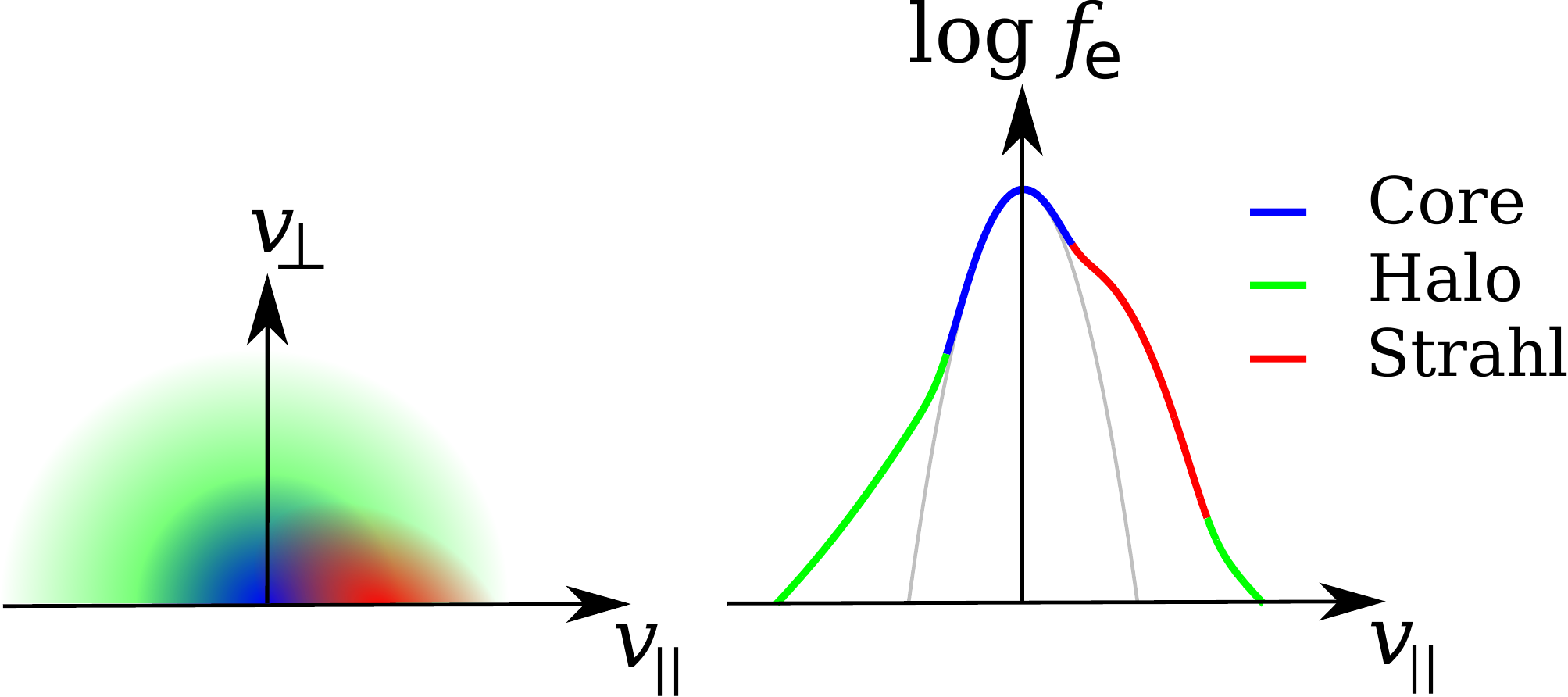}
\end{center}
\caption{ Schematic of a typical electron distribution function in the solar wind. Left: two-dimensional distribution function in $(v_{\perp},v_{\parallel})$ space. Right: cut of the distribution function along the $v_{\parallel}$-axis. The blue colour indicates the electron core, the green colour represents the halo, and the red colour represents the strahl. }\label{fig_schematic_VDF}
\end{figure}

Another important non-thermal feature of the solar-wind electron distribution function is its ternary structure consisting of a thermal core, a suprathermal halo \citep{feldman75, pilip87,lie-svendsen97,maksimovic97}, and a field-aligned beam \citep{pilip87,lin98}. Fig.~\ref{fig_schematic_VDF} illustrates the three populations of the electron distribution function in velocity space and the formation of the overall electron distribution in the solar wind.

The thermal core consists of about 95\% of the electrons (blue colour in Fig.~\ref{fig_schematic_VDF}). It has a shape close to a Maxwellian distribution and temperatures comparable to the proton temperatures in the solar wind. The Maxwellian shape of the core is often attributed to the lower mean free path for Coulomb collisions at low speeds in the distribution \citep{phillips90}.

The suprathermal halo is a quasi-isotropic tail of electrons represented by an enhancement of $f_{\mathrm e}$ above the Maxwellian distribution. It is primarily observed at energies above a breakpoint of about 50~eV at 1~au \citep[green colour in Fig.~\ref{fig_schematic_VDF};][]{mccomas92,lie-svendsen97}. The location of this breakpoint and the relative density of the halo population vary with distance from the Sun and show correlations with solar-wind parameters such as speed and temperature  \citep{maksimovic00,maksimovic05,Pierrard16,Pierrard20,bakrania20}. The halo population is often successfully modelled with a $\kappa$-distribution (using the Greek letter ``kappa''; for detailed information about $\kappa$-distributions, see the recent textbooks by \citealp{livadiotis17} and \citealp{lazar21}).

The field-aligned beam population is called the electron strahl (red colour in Fig.~\ref{fig_schematic_VDF}). This population appears as a ``shoulder'' on the electron distribution  at small pitch-angles around the directions parallel or anti-parallel to the magnetic field and typically in the anti-sunward direction \citep{hammond96,fitzenreiter98}. As in the case of the halo, the breakpoint energy between the core and the strahl populations and the relative density of the strahl vary with distance from the Sun and exhibit correlations with the solar-wind speed and temperature \citep{maksimovic05,pagel07,stverak09,graham17,abraham22}. The bulk velocities of the core, halo, and strahl often exhibit non-zero differences in their components parallel to the magnetic field. Given the requirement for global quasi-neutrality imposed by Poisson's equation, these field-aligned relative drifts must be such that the total electron charge flux is equal to the total ion charge flux. Given the outward drift of the strahl, this typically leads to a sunward drift of the core distribution. The relative drifts, particularly those of the suprathermal components, are responsible for the majority of the heat flux in the electron distribution.

If the deviations from thermodynamic equilibrium are large and certain criteria, which we discuss in this review, are fulfilled, the kinetic configuration of $f_{\mathrm e}$  drives kinetic micro-instabilities. These instabilities lead to the exponential growth of fluctuations in the electromagnetic or electrostatic fields over time at the expense of the integrated particle kinetic energy. During the growth of these instabilities, particles interact with the growing fluctuations, leading to a change of $f_{\mathrm e}$ that reduces the non-thermal drivers of the instability, until $f_{\mathrm e}$ achieves a marginally stable state. In the case of instabilities driven by temperature anisotropy, this process leads to a reduction of the anisotropy. In the case of instabilities driven by heat flux, this process leads to a reduction of the heat flux \citep{lopez20}. The efficiency of the heat-flux reduction by different instabilities is a matter of ongoing research. The ability of electron-driven instabilities to regulate electron temperatures, temperature anisotropies, and potentially heat flux makes them important for the overall evolution of the solar wind. We often characterise these instabilities in terms of instability thresholds that depend on plasma bulk parameters, such as the densities, bulk speeds, and temperatures of the involved plasma populations. 

The launch of \emph{Parker Solar Probe} in 2018 and the launch of \emph{Solar Orbiter} in 2020 have started a new era of electron observations in the solar wind \citep{fox16,mueller20,owen20,whittlesey20}. These spacecraft measure the three-dimensional solar-wind electron distribution function over a wide range of heliocentric distances and with unprecedented accuracy and cadence.  Electrons are particularly difficult to measure due to their small mass and due to the small kinetic energies of a large number of electrons in the distribution \citep{wuest07}. These energies are often comparable to the energy associated with the spacecraft electrostatic potential at the measurement point.  Nevertheless, these modern observations confirm earlier suggestions that the electron distribution evolves with distance from the Sun and that non-thermal features are essential for a complete description of the evolution of the solar wind, especially near the Sun \citep{halekas20,bercic21,halekas21,abraham22,jeong22a}. These results and extrapolations based on previous measurements also suggest that electron-driven instabilities play an important role in the shaping of the electron distribution \citep{bercic19}, although many questions about electron kinetics and its impact on the evolution of the solar wind remain open.

With this review, we pay tribute to the many theoretical and numerical discoveries made by Peter Gary in the field of electron-driven instabilities in the solar wind. Through his application of linear Vlasov--Maxwell theory, Peter made crucial contributions to the understanding of the energetics of the solar wind. In Section~\ref{theory}, we present a basic summary of linear Vlasov--Maxwell theory and our quasi-linear framework to visualise the impact of electron-driven instabilities. In Section~\ref{Inst_ani}, we discuss instabilities driven by temperature anisotropies in the solar-wind electron populations. In Section~\ref{Inst_asy}, we explore instabilities driven by reflectional asymmetries in the electron distribution function, including instabilities driven by electron heat flux. Section~\ref{Inst_nonres} gives a short summary of non-resonant electron-driven instabilities. Finally, Section~\ref{Conclusions} presents the conclusions of our work as well as an outlook on open questions and future observations of electron-driven instabilities.

\section{Theoretical framework for the description of resonant micro-instabilities}\label{theory}

In this section, we summarise linear Vlasov--Maxwell theory for the calculation of the hot-plasma dispersion relation of plasma waves and instabilities. We then introduce a quasi-linear framework for the description of the evolution of the electron distribution function under the action of electron-driven instabilities. The framework described in this section applies both to electrons and to ions in collisionless plasmas. This prepares us for the discussion in the subsequent sections of specific electron-driven instabilities in the solar wind.

\subsection{Linear Vlasov--Maxwell theory}

Linear Vlasov--Maxwell theory is a framework for the description of small-amplitude plasma waves in kinetic plasmas. The starting point for the derivation of the hot-plasma dispersion relation in linear Vlasov--Maxwell theory is the Vlasov equation,
\begin{equation}\label{Vlasov}
    \frac{\partial f_j}{\partial t}+\vec v\cdot \frac{\partial f_j}{\partial \vec x}+\frac{q_j}{m_j}\left(\vec E+\frac{1}{c}\vec v\times \vec B\right)\cdot\frac{\partial f_j}{\partial \vec v}=0,
\end{equation}
combined with Maxwell's equations,
\begin{equation}
    \nabla\cdot \vec E=4\pi\varrho,
\end{equation}
\begin{equation}
    \nabla \cdot \vec B=0,
\end{equation}
\begin{equation}
    \nabla \times \vec E=-\frac{1}{c}\frac{\partial \vec B}{\partial t},
\end{equation}
and
\begin{equation}\label{Ampere}
    \nabla \times \vec B=\frac{4\pi}{c}\vec j+\frac{1}{c}\frac{\partial \vec E}{\partial t}.
\end{equation}
In this coupled set of equations, $f_j(\vec x,\vec v,t)$ is the velocity distribution function of species $j$, $\vec E$ is the electric field, $\vec B$ is the magnetic field, $q_j$ and $m_j$ are the charge and the mass of a particle of species $j$, $\varrho$ is the charge density, $\vec j$ is the current density, and $c$ is the speed of light. Self-consistency demands that
\begin{equation}
    \varrho=\sum\limits_j q_j \int f_j\,\mathrm d^3v
\end{equation}
and
\begin{equation}
    \vec j=\sum\limits_j q_j \int \vec v f_j\,\mathrm d^3v,
\end{equation}
showing that Eqs.~(\ref{Vlasov}) through (\ref{Ampere}) represent a complicated, coupled set of integro-differential equations in six-dimensional phase space and time. Linear Vlasov--Maxwell theory simplifies this set of equations by linearisation so that 
\begin{equation}\label{fjdec}
f_j(\vec x,\vec v,t)=f_{0j}(\vec v)+\delta f_j(\vec x,\vec v,t), 
\end{equation}
$\vec E(\vec x,t)=\delta \vec E(\vec x,t)$, and $\vec B(\vec x,t)=\vec B_0+\delta \vec B(\vec x,t)$, where the subscript 0 indicates a \emph{background} quantity and $\delta$ indicates a \emph{fluctuating} quantity that averages to zero over time and space. Moreover, we make the assumption that all fluctuating quantities behave like plane waves, $\propto e^{i\vec k\cdot \vec x-\omega t}$, where $\vec k$ is the wave vector and $\omega$ is the wave frequency. As described in the literature \citep[e.g.,][]{stix92}, the application of these assumptions and Landau's procedure for the analytic continuation around poles in the complex plane lead to the dispersion relation in the form
\begin{equation}\label{dispersion}
    \text{det}\,\mathcal D=0,
\end{equation}
where 
\begin{equation}
\mathcal D=\begin{pmatrix}
\epsilon_{xx}-n_z^2 & \epsilon_{xy} & \epsilon_{xz}+n_xn_z\\
\epsilon_{yx} & \epsilon_{yy}-n_x^2-n_z^2 & \epsilon_{yz} \\
\epsilon_{zx} + n_zn_x & \epsilon_{zy} & \epsilon_{zz}-n_x^2
\end{pmatrix},
\end{equation}
is the dispersion tensor, $\vec \epsilon$ is the plasma susceptibility tensor, and $\vec n=\vec kc/\omega$. In this convention, the reference frame is chosen so that $k_y=0$. The entries of the $3\times 3$ matrix $\mathcal D$ depend on the plasma background properties ($q_j$, $m_j$, $f_{0j}$, and $\vec B_0$) and of the wave properties ($\vec k$ and $\omega$). Numerous numerical tools exist that solve Eq.~(\ref{dispersion}), often assuming closed expressions for $f_{0j}$ such as Maxwellian or bi-Maxwellian distributions \citep{roennmark82,klein12,verscharen18b}. 

The standard approach for finding the dispersion relation (corresponding to an initial-value problem) involves the determination of a complex $\omega$ that solves  Eq.~(\ref{dispersion}) for given plasma background properties at fixed $\vec k$. In general, these solutions are complex-valued. We define the complex $\omega$ that solves Eq.~(\ref{dispersion}) for given background parameters and $\vec k$ as
\begin{equation}\label{omegadec}
    \omega_k=\omega_{k\mathrm r}+i\gamma_k,
\end{equation}
where $\omega_{k\mathrm r}=\mathrm{Re}(\omega_k)$ is the real wave frequency and $\gamma_k=\mathrm{Im}(\omega_k)$ is the growth/damping rate at wave vector $\vec k$. The fluctuation amplitudes of solutions with $\gamma_k<0$ exponentially decrease with time, while the fluctuation amplitudes of solutions with $\gamma_k>0$ exponentially increase with time. Therefore, we refer to solutions with $\gamma_k<0$ as \emph{damped waves} and to solutions with $\gamma_k>0$ as \emph{instabilities}. 
In linear theory, the damping rate $\gamma_k$ is generally a function of $\vec k$ that possesses a global maximum at fixed plasma background properties. We refer to the maximum growth rate $\gamma_{\mathrm m}$ as the maximum $\gamma_k$ over all $\vec k$ for a given instability and given plasma background properties. 

Due to the kinetic (microphysical) nature of these instabilities, we also find the term \emph{micro-instabilities} for these solutions in the literature. Peter Gary pioneered the application of linear Vlasov--Maxwell theory to the study of micro-instabilities in space plasmas.

\subsection{Quasi-linear evolution of micro-instabilities}

If $\gamma_k\neq 0$, the energy density of the electromagnetic fluctuations changes over time.  This process exchanges energy between the electromagnetic field and the plasma particles, either in the form of a particle energy loss ($\gamma_k>0$) or gain ($\gamma_k<0$) in order to conserve the total energy. This fundamental concept helps us understand the evolution of the velocity distribution function under the action of micro-instabilities.

Resonant micro-instabilities are a family of micro-instabilities in which the energy exchange occurs via resonant wave--particle interactions between the unstable waves and the plasma particles. 

Quasi-linear theory is a mathematical framework to describe the evolution of $f_{0j}$ \citep{vedenov61,drummond64,kennel66a,rowlands66}.
It requires that the amplitude of the resonant waves and their damping rates are small ($|\delta f_j|\ll f_{0j}$ in Eq.~(\ref{fjdec}), and $|\gamma_{\mathrm m}|\ll|\omega_{k\mathrm r}|$ in Eq.~(\ref{omegadec}) at the unstable $\vec k$), so that 
the timescale of the evolution of $f_{0j}$ is much greater than the period of the resonant wave $1/\omega_{k\mathrm r}$.

Under the assumptions of quasi-linear theory, the background distribution is gyrotropic; i.e., cylindrically symmetric around $\vec B_0$. Therefore, it is helpful to work in cylindrical coordinates in velocity space, so that $\vec v$ is represented by the velocity component $v_{\perp}$ perpendicular to $\vec B_0$, the velocity component $v_{\parallel}$ parallel to $\vec B_0$, and the azimuthal angle $\phi_v$. Likewise, we express $\vec k$ with its cylindrical coordinates $k_{\perp}$, $k_{\parallel}$, and $\phi_k$.

The slow, quasi-linear evolution of $f_{0j}$ over time due to resonant wave--particle interactions is given by the equation \citep{stix92}
\begin{equation}\label{QLdiff}
    \frac{\partial f_{0j}}{\partial t}=\lim\limits_{V\rightarrow \infty} \sum\limits_{n=-\infty}^{+\infty} \frac{q_j^2}{8\pi^2m_j^2}\int \frac{1}{v_{\perp}V}\hat G v_{\perp} \delta\left(\omega_{k\mathrm r}-k_{\parallel}v_{\parallel}-n\Omega_j\right)\left|\psi_k^{j,n}\right|^2\hat G f_{0j}\,\mathrm d^3k,
\end{equation}
where
\begin{equation}
    \hat G\equiv \left(1-\frac{k_{\parallel}v_{\parallel}}{\omega_{k\mathrm r}}\right)\frac{\partial }{\partial v_{\perp}}+\frac{k_{\parallel}v_{\perp}}{\omega_{k\mathrm r}}\frac{\partial }{\partial v_{\parallel}},
\end{equation}
\begin{equation}\label{psi}
    \psi_k^{j,n}\equiv \frac{1}{\sqrt{2}}\left[E_{k,\mathrm r}e^{i\phi_k}J_{n+1}(\xi_j)+E_{k,\mathrm l}e^{-i\phi_k}J_{n-1}(\xi_j)\right]+\frac{v_{\parallel}}{v_{\perp}}E_{kz}J_n(\xi_j),
\end{equation}
$\Omega_j\equiv q_jB_0/m_jc$ is the cyclotron frequency\footnote{In our convention, $\Omega_j$ has the same sign as $q_j$. This means particularly with regard to electrons that $\Omega_{\mathrm e}<0$.}, $\xi_j=k_{\perp}v_{\perp}/\Omega_j$ is the argument of the Bessel function $J_m$ of order $m$, and $n$ is an integer that marks the order of the resonance. We refer to the resonance with $n=0$ as the \emph{Landau resonance} and to all other resonances with $n\neq 0$ as \emph{cyclotron resonances}. The left and right circularly polarised components of the electric field are given by $E_{k,\mathrm l}\equiv \left(E_{kx}+iE_{ky}\right)/\sqrt{2}$ and $E_{k,\mathrm r}\equiv \left(E_{kx}-iE_{ky}\right)/\sqrt{2}$, where we use the Fourier transformation of the electric field in the convention
\begin{equation}\label{EkF}
    \vec E_k(\vec k,t)=\int_V\delta \vec E(\vec x,t)e^{-i\vec k\cdot \vec x}\,\mathrm d^3x
\end{equation}
over the spatial volume $V$. We define the sense of the polarisation of a given wave mode in terms of $E_{k,\mathrm l}$, $E_{k,\mathrm r}$, and $E_{kz}$.

Due to the $\delta$-function in Eq.~(\ref{QLdiff}), only particles fulfilling the resonance condition
\begin{equation}\label{rescond}
    \omega_{k\mathrm r}=k_{\parallel}v_{\parallel}+n\Omega_j
\end{equation}
participate in the resonant wave--particle interactions associated with a given $n$. In the case of Landau-resonant interactions, a resonant particle travels along $\vec B_0$ with the parallel phase speed of the resonant wave, $v_{\parallel}=\omega_{k\mathrm r}/k_{\parallel}$. This resonant particle experiences a constant parallel wave electric field $E_{\parallel}\equiv \delta \vec E\cdot \vec B_0/B_0$. In the case of cyclotron-resonant interactions, a resonant particle travels along $\vec B_0$ at a speed such that the Doppler-shifted wave frequency in the particle's frame of reference is an integer multiple of the particle's cyclotron frequency. Such a particle experiences a perpendicular wave electric field $\vec E_{\perp}\equiv \delta \vec E-E_{\parallel}\vec B_0/B_0$ that does not average to zero over multiple cyclotron periods of the particle. This description captures the fundamental nature of resonant wave--particle interactions. We note, however, that Eq.~(\ref{QLdiff}) includes more subtleties such as higher-order and anomalous cyclotron resonances as well as transit-time damping, which lie outside the scope of this review.

\begin{figure}[ht]
\begin{center}
\includegraphics[width=\textwidth]{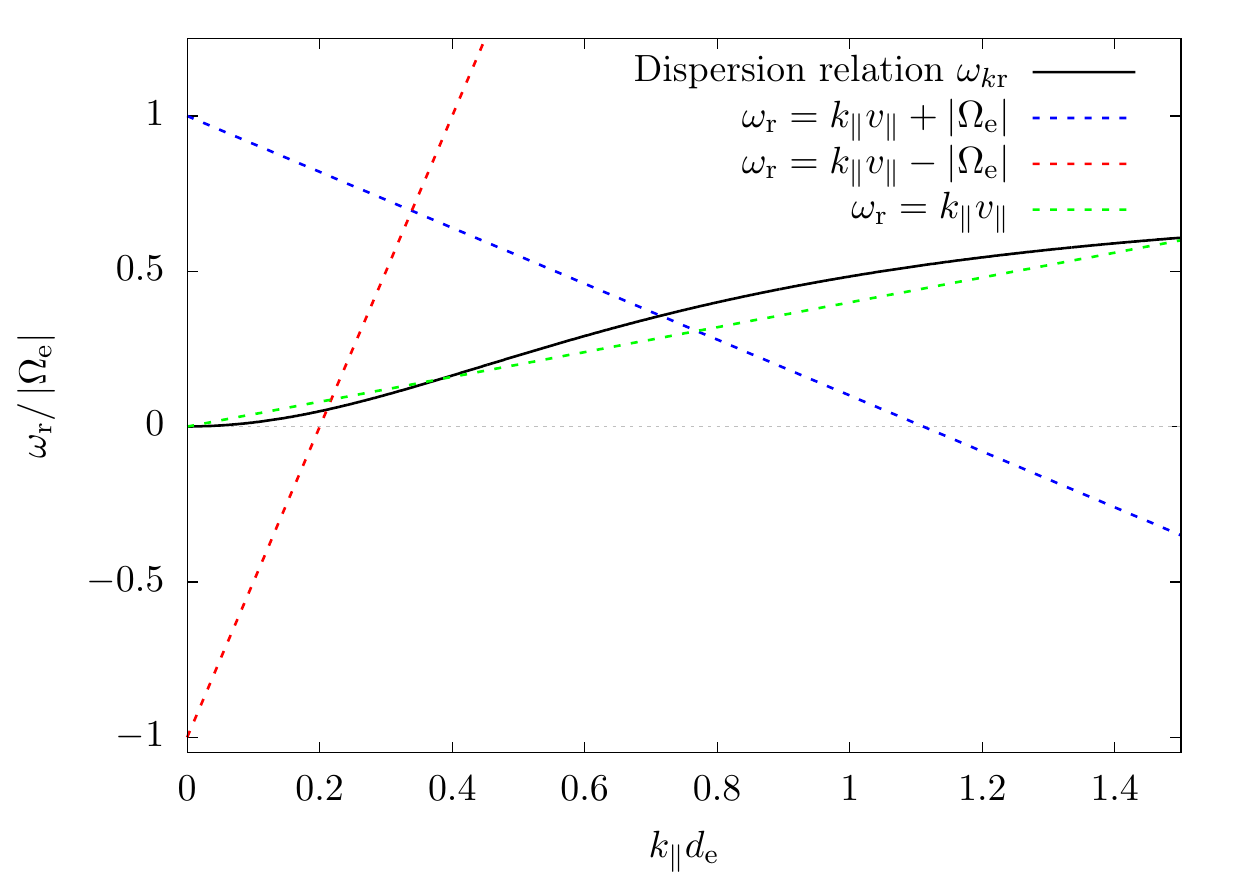}
\end{center}
\caption{ Dispersion relation and resonance conditions. The black curve shows a solution to Eq.~(\ref{dispersion}) for the fast-magnetosonic/whistler wave. We assume an angle of 40$^{\circ}$ between $\vec k$ and $\vec B_0$ and a very cold plasma ($\beta_j=10^{-8}$) consisting of Maxwellian protons and electrons. For our definition of $\beta_j$, see Eq.~(\ref{beta}). The dashed lines represent electron resonance conditions according to Eq.~(\ref{rescond}) for different $n$ and $v_{\parallel}$. }\label{fig_dispersion_analysis}
\end{figure}
Figure~\ref{fig_dispersion_analysis} illustrates an example for a solution to the linear Vlasov--Maxwell dispersion relation from Eq.~(\ref{dispersion}) and the resonance conditions from quasi-linear theory in Eq.~(\ref{rescond}). The black curve shows a solution of the Vlasov--Maxwell dispersion relation from Eq.~(\ref{dispersion}) in terms of $\omega_{k\mathrm r}$ for the fast-magnetosonic/whistler wave as a function of $k_{\parallel}$. The dashed lines describe the resonance conditions from Eq.~(\ref{rescond}) for $n=-1$ (blue), $n=+1$ (red), and $n=0$ (green). Their slopes correspond to different values of $v_{\parallel}$.  At any intersection between a line representing a resonance condition and the plot of the dispersion relation, Eq.~(\ref{rescond}) is fulfilled. 

Eq.~(\ref{QLdiff}) represents a diffusion equation in velocity space. The operator $\hat G$ dictates the direction of the diffusion when the resonance condition is fulfilled and $\psi_k^{j,n}\neq 0$. The diffusive flux of resonant particles through velocity space is locally tangent to semicircles in velocity space of the form
\begin{equation}\label{circle}
    \left(v_{\parallel }-v_{\mathrm{ph}}\right)^2+v_{\perp}^2=\text{constant},
\end{equation}
where $v_{\mathrm{ph}}\equiv\omega_{k{\mathrm r}}/k_{\parallel}$ is the field-parallel phase speed of the resonant waves. According to Eq.~(\ref{circle}), quasi-linear diffusion conserves particle kinetic energy in the reference frame that moves with the velocity $v_{\mathrm{ph}}\vec B_0/B_0$. The description of Eq.~(\ref{circle}) leaves us with an ambiguity in the direction of the diffusive flux of resonant particles (clockwise or counter-clockwise in velocity space). This ambiguity is resolved by the requirement that Eq.~(\ref{QLdiff}) demands, like other diffusion processes, a diffusive flux from larger values of $f_{0j}$ to smaller values of $f_{0j}$. 

Our cylindrical coordinate system is aligned with $\vec B_0$, so that waves with $v_{\mathrm{ph}}>0$ propagate in the direction of $\vec B_0$. The direction of propagation of wave solutions with respect to $\vec B_0$ can be reversed mathematically in two ways: either by changing the sign of $\omega_{k\mathrm r}$ or by changing the sign of $k_{\parallel}$.  Although this choice does not affect the physics described by the wave theory, it has an impact on the polarisation and thus the applicable resonance condition. We implicitly assume that   $\omega_{k\mathrm r}\ge 0$ throughout this work, so that a reversal of the direction of propagation corresponds to changing the sign of $k_{\parallel}$ in our convention\footnote{In this convention, $E_{k,\mathrm l}$ corresponds to left-hand polarisation and $E_{k,\mathrm r}$ corresponds to right-hand polarisation. The meaning of $E_{k,\mathrm l}$ and $E_{k,\mathrm r}$ for the characterisation of the sense of polarisation as left-circular and right-circular swaps when $\omega_{k\mathrm r}<0$.
In a plasma with symmetric distribution functions around $v_{\parallel}=0$, a forward-propagating wave solution with $\omega_{k\mathrm r}>0$ and $k_{\parallel}>0$ has the same sense of polarisation in terms of $E_{k,\mathrm l}$ and $E_{k,\mathrm r}$ as the corresponding backward-propagating solution with $\omega_{k\mathrm r}>0$ and $k_{\parallel}<0$.}.

\begin{figure}[ht]
\begin{center}
\includegraphics[width=0.6\textwidth]{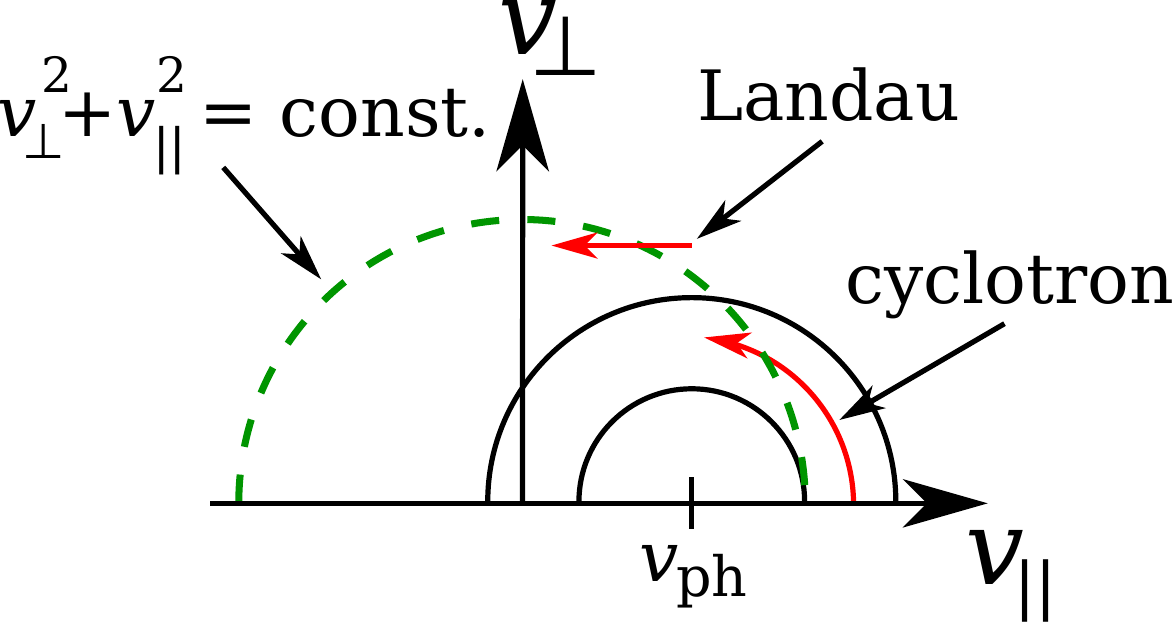}
\end{center}
\caption{Schematic illustration of quasi-linear diffusion in velocity space. The black semi-circles represent Eq.~(\ref{circle}) for a given parallel phase speed $v_{\mathrm{ph}}$. The diffusive flux of resonant particles is locally tangent to these semi-circles (marked by the red arrows). The green-dashed semi-circle indicates $v_{\perp}^2+v_{\parallel}^2=\text{constant}$. If the diffusive flux crosses the green-dashed semi-circle from larger to smaller values of $(v_{\perp}^2+v_{\parallel}^2)$, the resonant wave--particle interaction contributes to the growth of the resonant waves with parallel phase speed $v_{\mathrm{ph}}$. Examples for Landau-resonant and cyclotron-resonant interactions are indicated.}\label{fig_schematic_general}
\end{figure}
Figure~\ref{fig_schematic_general} illustrates the quasi-linear diffusion in velocity space. The black semi-circles represent Eq.~(\ref{circle}). Particles at $v_{\parallel}=v_{\mathrm{ph}}$ fulfill the Landau-resonance condition  with $n=0$ according to Eq.~(\ref{rescond}).  If $\hat Gf_{0j}>0$ at $v_{\parallel}=v_{\mathrm{ph}}$, the direction of the diffusive flux of Landau-resonant particles is as indicated by the red arrow marked ``Landau''. For Landau-resonant particles, $\hat Gf_{0j}=(v_{\perp}/v_{\mathrm{ph}})(\partial f_{0j}/\partial v_{\parallel})$. 

Assuming that particles with $v_{\parallel}=v_{\mathrm{res}} > v_{\mathrm{ph}}$ fulfill a cyclotron resonance condition according to Eq.~(\ref{rescond}) in this example, the diffusive flux of these cyclotron-resonant particles is as indicated by the red arrow marked ``cyclotron'' as long as $\hat Gf_{0j}>0$ at $v_{\parallel}=v_{\mathrm{res}}$.

As the resonant particles diffuse through velocity space, their $v_{\parallel}$ changes. If the particles  interact with waves with $k_\perp=0$ on only a single branch of the dispersion relation, then at each value of $v_\parallel$ the particles typically resonate with waves at a single value of $k_\parallel$. This resonant value of $k_\parallel$, which can be regarded as a function of $v_\parallel$, corresponds to a  unique parallel phase velocity $v_{\mathrm{ph}}$, and
over time the particles trace out a one-dimensional diffusion contour in velocity space that is locally tangent to the semicircles defined by Eq.~(\ref{circle}). For dispersive waves, $v_{\mathrm{ph}}$ varies with $k_\parallel$, and hence with $v_\parallel$ of the resonant particles, and thus the full diffusion contour is not semi-circular, because the centre of the locally tangent semi-circle  evolves during the diffusion process.

The quasi-linear evolution according to the concepts outlined so far generally leads to a change of the integrated particle kinetic energy of $f_{0j}$ (i.e., the second velocity moment of $f_{0j}$). If the kinetic energy $m_j(v_{\perp}^2+v_{\parallel}^2)/2$ of the resonant particles decreases in the quasi-linear diffusion process, this energy is transferred into the resonant waves, leading to growth of the wave amplitude and thus instability. If the energy of the resonant particles increases, this energy is taken from the resonant waves, corresponding to wave damping. Our graphical representation in Fig.~\ref{fig_schematic_general} allows us to evaluate the energy behaviour by comparing the direction of the diffusive flux of resonant particles with semi-circles around the origin (green-dashed in Fig.~\ref{fig_schematic_general}). These semi-circles represent isocontours of constant $(v_{\perp}^2+v_{\parallel}^2)$. If the direction of the diffusive flux locally crosses these semi-circles from larger to smaller $(v_{\perp}^2+v_{\parallel}^2)$, the process corresponds to an instability. If it crosses these semi-circles in the other direction, the process corresponds to damping. In the specific example shown in Fig.~\ref{fig_schematic_general}, both the marked Landau-resonant and the marked cyclotron-resonant particles contribute to an instability of the resonant wave at phase speed $v_{\mathrm{ph}}$. The question of damping/instability thus ultimately simplifies to an investigation of the relative alignments between the semi-circles in velocity space defined by Eq.~(\ref{circle}), the isocontours of $f_{0j}$, and the isocontours of $(v_{\perp}^2+v_{\parallel}^2)$ at the speed that fulfills Eq.~(\ref{rescond}) for resonant wave--particle interactions.

The propagation direction and the polarisation of the waves under consideration have a strong impact on the quasi-linear diffusion process. If the waves are purely parallel-propagating (i.e., $k_{\perp}=0$), then $\xi_j=k_{\perp}v_{\perp}/\Omega_j=0$ in  Eq.~(\ref{psi}). The Bessel functions have the property
\begin{equation}
    J_m(0)=\begin{cases}
			1 & \text{if } m=0,\\
            0 & \text{otherwise}.
		 \end{cases}
\end{equation}
This property simplifies Eq.~(\ref{QLdiff}) considerably for parallel-propagating waves. According to Eq.~(\ref{psi}), parallel-propagating waves only have $\psi_k^{j,n}\neq 0$ if $n=+1$, $n=-1$, or $n=0$.  If a parallel-propagating wave is purely left-circularly polarised (i.e., if $E_{k,\mathrm r}=E_{kz}=0$), only the cyclotron resonance with $n=+1$ contributes to the sum in Eq.~(\ref{QLdiff}). If a parallel-propagating wave is purely right-circularly polarised  (i.e., if $E_{k,\mathrm l}=E_{kz}=0$), only the cyclotron resonance with $n=-1$ contributes to the sum in Eq.~(\ref{QLdiff}). Lastly, if a parallel-propagating wave is purely longitudinal  (i.e., if $E_{k,\mathrm r}=E_{k,\mathrm l}=0$), only the Landau resonance with $n=0$ contributes to the sum in Eq.~(\ref{QLdiff}). This simplification of the quasi-linear diffusion equations is particularly useful since many instabilities have maximum growth for $k_{\perp}=0$, in which case they exhibit these pure polarisation properties according to linear Vlasov--Maxwell theory. For example, the parallel-propagating fast-magnetosonic/whistler wave is purely right-circularly polarised. The parallel-propagating Alfv\'en/ion-cyclotron wave is purely left-circularly polarised, and the parallel-propagating Langmuir wave is purely longitudinal.   We note that these definitions only apply when $\omega_{k\mathrm r}>0$ according to our convention. 
For waves with oblique wave vectors, in general, all $n$ must be considered and the polarisation is typically ``mixed'' with contributions from non-zero $E_{k,\mathrm r}$, $E_{k,\mathrm l}$, and $E_{kz}$. Nevertheless, it is often useful to consider that the Bessel-function contributions $J_m(\xi_j)$ in Eq.~(\ref{QLdiff}) are greater for $m=0$ than for other $m$ when $\xi_j$ is moderately small, which is often the case for the majority of the resonant particles.
 
In addition to its mathematical rigour, the quasi-linear-diffusion framework provides us with a visual aid to understand the physics of resonant micro-instabilities. 
It can be shown that the description of wave damping and instability in terms of quasi-linear diffusion is consistent with its description in terms of $\gamma_k$ from solutions of linear Vlasov--Maxwell theory as far as the assumptions of both frameworks are fulfilled (\citealp{kennel66a}; \citealp{kennel67}; see also \citealp{chandran10}).

The instabilities discussed in this review  occur on a variety of length scales, which are often related to the characteristic plasma scales of the system \citep{verscharen19a}. We define the inertial length of species $j$ as 
\begin{equation}
    d_j=\frac{c}{\omega_{\mathrm pj}}=\frac{v_{\mathrm Aj}}{\left|\Omega_j\right|}=\sqrt{\frac{m_jc^2}{4\pi n_jq_j^2}},
\end{equation}
where $v_{\mathrm Aj}\equiv B_0/\sqrt{4\pi n_jm_j}$ is the Alfv\'en speed of species $j$ and $n_j$ is the background number density of species $j$.
We define the gyro-radius of species $j$ as
\begin{equation}
    \rho_j=\frac{w_{\perp j}}{\left|\Omega_j\right |}=\sqrt{\frac{2k_{\mathrm B}T_{\perp j}m_jc^2}{q_j^2B_0^2}},
\end{equation}
where $w_{\perp j}\equiv \sqrt{2k_{\mathrm B}T_{\perp j}/m_j}$ is the perpendicular thermal speed of species $j$ and $k_{\mathrm B}$ is the Boltzmann constant.
For electrostatic instabilities, the Debye length
\begin{equation}
    \lambda_j=\sqrt{\frac{k_{\mathrm B}T_{\parallel j}}{4\pi n_jq_j^2}}
\end{equation}
of species $j$ defines an important spatial reference scale.
Lastly, we define the following dimensionless ratios of  kinetic to magnetic pressure:
\begin{equation}\label{beta}
    \beta_{j}\equiv \frac{8\pi n_j k_{\mathrm B}T_j}{B_0^2},\qquad \beta_{\perp j}\equiv \frac{8\pi n_j k_{\mathrm B}T_{\perp j}}{B_0^2}, \qquad \text{and}\qquad \beta_{\parallel j}\equiv \frac{8\pi n_j k_{\mathrm B}T_{\parallel j}}{B_0^2},
\end{equation}
where $T_j$ is the scalar temperature of species $j$, which we use in the case of isotropic plasmas when $T_{j}=T_{\perp j}=T_{\parallel j}$.

\section{Resonant instabilities driven by temperature anisotropies}\label{Inst_ani}

We discuss two types of resonant instabilities driven by electron temperature anisotropy: the electron whistler anisotropy instabilty and the propagating electron firehose instability. Both instabilities are associated with electromagnetic normal modes of the plasma. Under typical solar-wind conditions, non-resonant anisotropy-driven instabilities often have lower thresholds than the resonant instabilities. The non-resonant instabilities are discussed in Section~\ref{Inst_nonres}. In plasmas with $\omega_{\mathrm{pe}}<|\Omega_{\mathrm e}|$, electrostatic instabilities exist that are driven by electron anisotropy \citep{gary99a}. However, since this condition is not fulfilled in the solar wind, we do not discuss these instabilities further.

\subsection{Electron whistler anisotropy instability}\label{sec_whistler_ani}

\begin{figure}[ht]
\begin{center}
\includegraphics[width=0.6\textwidth]{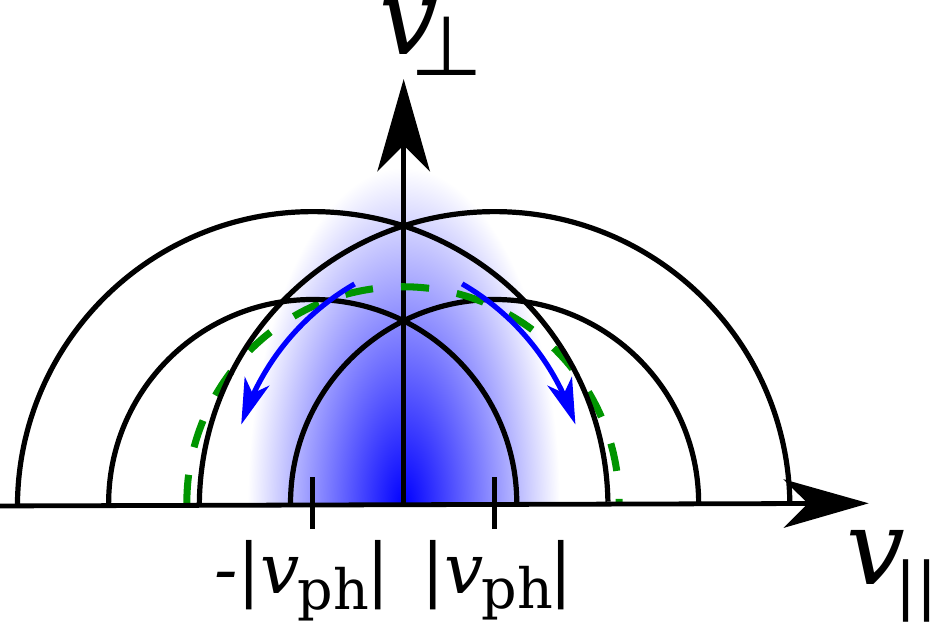}
\end{center}
\caption{
Schematic illustration of the quasi-linear diffusion in the  electron whistler anisotropy instability. The blue shaded area represents the anisotropic electron population with $T_{\perp\mathrm e}>T_{\parallel\mathrm e}$. The black semi-circles represent Eq.~(\ref{circle}) for fast-magnetosonic/whistler waves with parallel phase speed $|v_{\mathrm{ph}}|$ (propagating in the direction parallel to $\vec B_0$) and $-|v_{\mathrm{ph}}|$ (propagating in the direction anti-parallel to $\vec B_0$). The diffusive flux of cyclotron-resonant particles is shown by the blue arrows. The green-dashed semi-circle indicates $v_{\perp}^2+v_{\parallel}^2=\text{constant}$.
}\label{fig_schematic_whistler_temp}
\end{figure}

The electron whistler anisotropy instability is driven when $T_{\perp\mathrm e}>T_{\parallel\mathrm e}$ \citep{kennel66,scharer67,gary06,lazar22}. It is an instability of the fast-magnetosonic/whistler  wave with $\Omega_{\mathrm p}\ll \omega_{k\mathrm r}< |\Omega_{\mathrm e}|$ and  $k\lesssim 1/d_{\mathrm e}$ at maximum growth. The instability has maximum growth when $k_{\perp}=0$.

Fig.~\ref{fig_schematic_whistler_temp} describes the quasi-linear evolution of $f_{0\mathrm e}$ under the action of the electron whistler anisotropy instability. The initial electron distribution is elongated (i.e., anisotropic) in the direction perpendicular to the background magnetic field. Electrons with $v_{\parallel}>0$ resonate with fast-magnetosonic/whistler waves at parallel phase speed $-|v_{\mathrm{ph}}|$ (i.e., propagating oppositely to the direction of $\vec B_0$) through the $n=+1$ cyclotron resonance according to Eq.~(\ref{rescond}). Electrons with $v_{\parallel}<0$ resonate with waves at parallel phase speed $+|v_{\mathrm{ph}}|$ through the $n=-1$ cyclotron resonance according to Eq.~(\ref{rescond}). 

The relative alignment between the isocontours of $f_{0\mathrm e}$ at the value of $v_{\parallel}$ that fulfills Eq.~(\ref{rescond}) and the black semi-circles around $\pm |v_{\mathrm{ph}}|$ in Fig.~\ref{fig_schematic_whistler_temp} guarantees that the quasi-linear diffusion is locally directed tangent to the blue arrows. Therefore, $v_{\perp}$ of the resonant electrons decreases while their $|v_{\parallel}|$ increases. Overall, this process leads to a reduction of $(v_{\perp}^2+v_{\parallel}^2)$ and thus of the energy of the resonant electrons. This energy is transferred into the resonant fast-magnetosonic/whistler waves in both directions of propagation. As in the case of the propagating electron firehose instability presented in Section~\ref{par_firehose}, the overall temperature anisotropy of the distribution decreases in this process until the distribution function relaxes to a quasi-linear plateau in the part of velocity space occupied by resonant electrons. If $f_{0\mathrm e}$ is asymmetric around $v_{\parallel}=0$, the energy in the unstable counter-propagating waves can be imbalanced.

 \citet{kennel66} give a necessary condition for instability of the electron whistler anisotropy instability as
\begin{equation}
    \frac{T_{\perp\mathrm e}}{T_{\parallel\mathrm e}}-1>\frac{1}{\frac{|\Omega_{\mathrm e}|}{\omega_{k\mathrm r}}-1}.
\end{equation}
The unstable mode follows the approximate dispersion relation  \citep{gary93}
\begin{equation}
    \frac{\omega_{k\mathrm r}}{\Omega_{\mathrm p}}\simeq k_{\parallel}^2d_{\mathrm p}^2\left[1+\left(\frac{T_{\perp\mathrm e}}{T_{\parallel\mathrm e}}-1\right)\frac{\beta_{\parallel \mathrm e}}{2}\right].
\end{equation}
When $T_{\parallel \mathrm e}\sim T_{\parallel \mathrm p}$ and $\beta_{\parallel \mathrm e}\sim 1$ as in the solar wind, protons are unlikely to undergo a significant resonant interaction with parallel-propagating fast-magnetosonic/whistler waves. Therefore, this instability does not compete with proton-resonant damping. 

The necessary relative alignment between the isocontours of $f_{0\mathrm e}$ at the value of $v_{\parallel}$ that fulfills Eq.~(\ref{rescond}) and the semi-circles around $\pm |v_{\mathrm{ph}}|$ as shown in Fig.~\ref{fig_schematic_whistler_temp} can also be fulfilled in bi-$\kappa$ electron distributions \citep{lazar11,lazar13,shaaban21}. Likewise, the instability criteria can also be fulfilled in plasmas consisting of anisotropic core and halo populations \citep{gary12,lazar18b}. In these cases, anisotropic halo electrons resonate with parallel-propagating fast-magnetosonic/whistler waves through the same mechanism as the electron core \citep{lazar15}. If the core is isotropic, the halo driving competes with cyclotron-resonant core damping.

The electron whistler anisotropy instability is believed to be responsible for the sporadic generation of parallel-propagating fast-magnetosonic/whistler waves that are intermittently observed in the solar wind \citep{tong19a,jagarlamudi20,vasko20}. Observations show that the solar wind very rarely exhibits plasma conditions above the instability threshold though \citep{stverak08}.

A review of early simulation work of the electron whistler anisotropy instability with quasi-linear context is given by \citet{cuperman81}. Particle-in-cell simulations reveal that this instability changes its properties in the low-$\beta_{\parallel \mathrm e}$ regime, in which the wave at maximum growth is predominantly oblique and electrostatic, so that Landau-resonant processes become important \citep{gary11}. Kinetic simulations of the  electron whistler anisotropy instability agree reasonably well with quasi-linear predictions in terms of the behaviour of the instability at saturation \citep{kim2017}. In particular, these simulations indicate the occurrence of weakly resonant wave--particle interactions. While these numerical simulations start with bi-Maxwellian electron distributions, more recently, the impact of more realistic electron distribution functions has been explored. For instance, \citet{lazar22} perform particle-in-cell simulations with bi-$\kappa$ electron distributions with different $\beta_{\parallel \mathrm e}$. In the explored cases,  the presence of suprathermal electrons leads to higher growth rates and oscillation amplitudes than in the bi-Maxwellian case.

\subsection{Propagating electron firehose instability}\label{par_firehose}

\begin{figure}[ht]
\begin{center}
\includegraphics[width=0.6\textwidth]{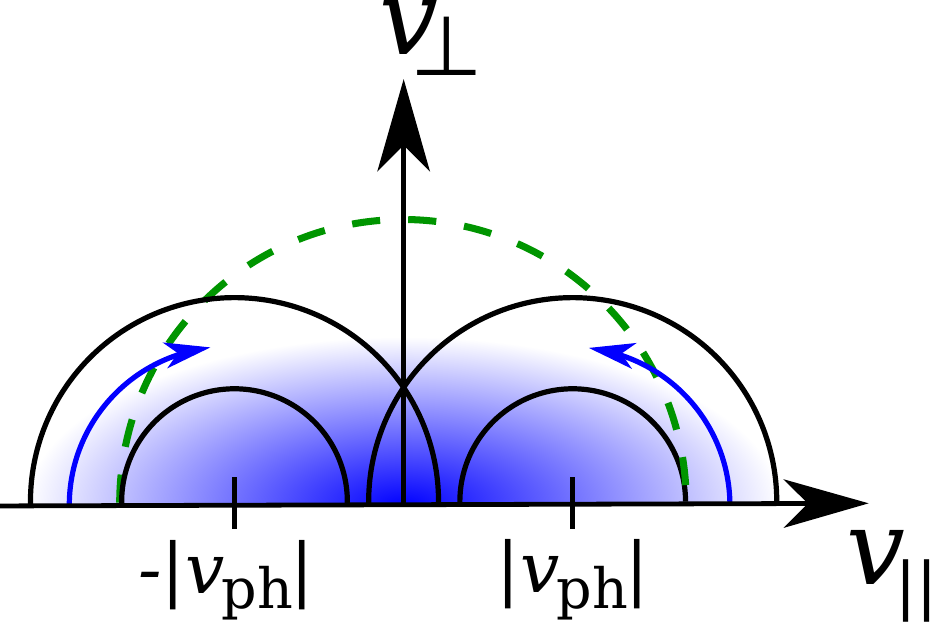}
\end{center}
\caption{
Schematic illustration of the quasi-linear diffusion in the propagating electron firehose instability. The blue shaded area represents the anisotropic electron population with $T_{\parallel\mathrm e}>T_{\perp\mathrm e}$. The black semi-circles represent Eq.~(\ref{circle}) for modified fast-magnetosonic/whistler waves with parallel phase speed $|v_{\mathrm{ph}}|$ (propagating in the direction parallel to $\vec B_0$) and $-|v_{\mathrm{ph}}|$ (propagating in the direction anti-parallel to $\vec B_0$). The diffusive flux of cyclotron-resonant particles is shown by the blue arrows. The green-dashed semi-circle indicates $v_{\perp}^2+v_{\parallel}^2=\text{constant}$.
}\label{fig_schematic_firehose}
\end{figure}

The propagating electron firehose instability is driven when $T_{\parallel\mathrm e}>T_{\perp\mathrm e}$ \citep{hollweg70,pilipp71,li00}. 
It corresponds to an instability of left-hand polarised fast-magnetosonic/whistler modes that undergo a significant change in their dispersion relation under the relevant unstable plasma conditions. In the following discussion, we focus on the parallel-propagating case, in which $k_\perp=0$.

When $T_{\perp \mathrm e} = T_{\parallel \mathrm e}$,
the fast-magnetosonic/whistler branch of the dispersion relation is right-circularly polarised. However, when $T_{\perp \mathrm e}/T_{\parallel \mathrm e}$ is sufficiently small, the fast-magnetosonic/whistler wave becomes left-circularly polarised as $|k_\parallel|$ increases from small to large values. These left-circularly polarised fast-magnetosonic/whistler waves can interact with electrons when Eq.~(\ref{rescond}) is satisfied for the $n=+1$ resonance, and such interactions can drive the wave unstable.  When unstable, this mode satisfies $\Omega_{\mathrm p}<\omega_{k\mathrm r}\ll |\Omega_{\mathrm e}|$ and $1/d_{\mathrm p}<k_{\parallel}<1/d_{\mathrm e}$ \citep{micera2020particle}.

Fig.~\ref{fig_schematic_firehose} describes the quasi-linear evolution of $f_{0\mathrm e}$ under the action of the propagating electron firehose instability. The initial electron distribution is elongated (i.e., anisotropic) in the direction parallel to the background magnetic field. Electrons with $v_{\parallel}>|v_{\mathrm{ph}}|$ resonate with waves at $+|v_{\mathrm{ph}}|$ through the $n=+1$ cyclotron resonance according to Eq.~(\ref{rescond}). Electrons with $v_{\parallel}<-|v_{\mathrm{ph}}|$ resonate with waves propagating in the anti-parallel direction to $\vec B_0$ (i.e., with a phase speed $-|v_{\mathrm{ph}}|$) through the $n=-1$ cyclotron resonance according to Eq.~(\ref{rescond}).

The relative alignment between the isocontours of $f_{0\mathrm e}$ at the value of $v_{\parallel}$ that fulfills Eq.~(\ref{rescond}) and the black semi-circles around $\pm |v_{\mathrm{ph}}|$ in Fig.~\ref{fig_schematic_firehose} guarantees that the quasi-linear diffusion is locally directed tangent to the blue arrows. Therefore, $v_{\perp}$ of the resonant electrons increases while their $|v_{\parallel}|$ decreases. Overall,  this process leads to a reduction of $(v_{\perp}^2+v_{\parallel}^2)$ and thus of the kinetic energy of the resonant electrons. This energy is transferred into the resonant fast-magnetosonic/whistler waves in both directions of propagation. As in the case of the electron whistler anisotropy instability discussed in Section~\ref{sec_whistler_ani}, the overall temperature anisotropy of the distribution decreases in this process until the distribution function relaxes to a quasi-linear plateau in the part of velocity space occupied by resonant electrons. If $f_{0\mathrm e}$ is asymmetric around $v_{\parallel}=0$, the energy in the unstable counter-propagating waves can be imbalanced. 

In addition to anisotropic core electrons, also suprathermal electron populations such as an anisotropic halo with a bi-$\kappa$ distribution can drive the propagating electron firehose instability \citep{lazar17,shaaban21}.

Since this instability is left-hand polarised, cyclotron-resonant proton damping counteracts the driving by cyclotron-resonant electrons. Due to its high frequency compared to $\Omega_{\mathrm p}$, the instability's growth rate depends only weakly on $T_{\parallel \mathrm e}/T_{\parallel \mathrm p}$ and  $T_{\perp \mathrm p}/T_{\parallel \mathrm p}$ \citep{hollweg70,gary85a}.
The non-propagating firehose instability discussed in Section~\ref{oblFH} often has a lower threshold than the propagating firehose instability under most solar-wind conditions \citep{paesold99,li00,gary03}.

One-dimensional, relativistic particle-in-cell simulations of the propagating electron firehose instability underline its possible role as a temperature-isotropisation mechanism in solar-flare plasmas \citep{paesold99,messmer2002temperature}. Simulations with both anisotropic protons and electrons reveal that the concurrent presence of a proton and electron temperature anisotropy can increase the growth rate of the propagating \emph{proton} firehose instability compared to plasmas with isotropic electrons \citep{micera2020particle}.
Quasi-parallel and exactly parallel electron firehose modes  become dominant after the saturation of oblique modes with higher growth rates  \citep[see also Section~\ref{oblFH};][the latter study is conducted within an expanding-box framework]{camporeale2008electron,innocenti2019onset}.

\section{Resonant instabilities driven by reflectional asymmetries in the distribution function}\label{Inst_asy}

In this section, we discuss instabilities driven by asymmetries in the electron distribution function around $v_{\parallel}=0$. These asymmetries can be represented by beams, multi-beam structures, or skewness in the electron distribution \citep{forslund70}. We distinguish between electrostatic and electromagnetic instabilities driven by reflectional asymmetries in the distribution function. The electrostatic approximation is valid in plasmas with $\beta_j\ll 1$ for all $j$. In this case, $\vec E\approx -\nabla \phi$, where $\phi$ is the electrostatic potential, and $\vec B\approx \vec B_0$.
With increasing $\beta_j$, however, the coupling between electric and magnetic fluctuations increases, and the fluctuations become increasingly electromagnetic. Nevertheless, some electrostatic modes also exist in plasmas with moderate to high $\beta_j$, especially when they propagate along $\vec B_0$.
Electromagnetic beam instabilities compete with their electrostatic counterparts in the presence of hot electron beams and reasonably large $\beta_j$, which is often (but not always) the case in the solar wind. 
Unless stated otherwise, we work in the reference frame in which the background bulk speed of the protons is zero.

\subsection{Electron/ion-acoustic instability}\label{sec_iaw}

\begin{figure}[ht]
\begin{center}
\includegraphics[width=0.6\textwidth]{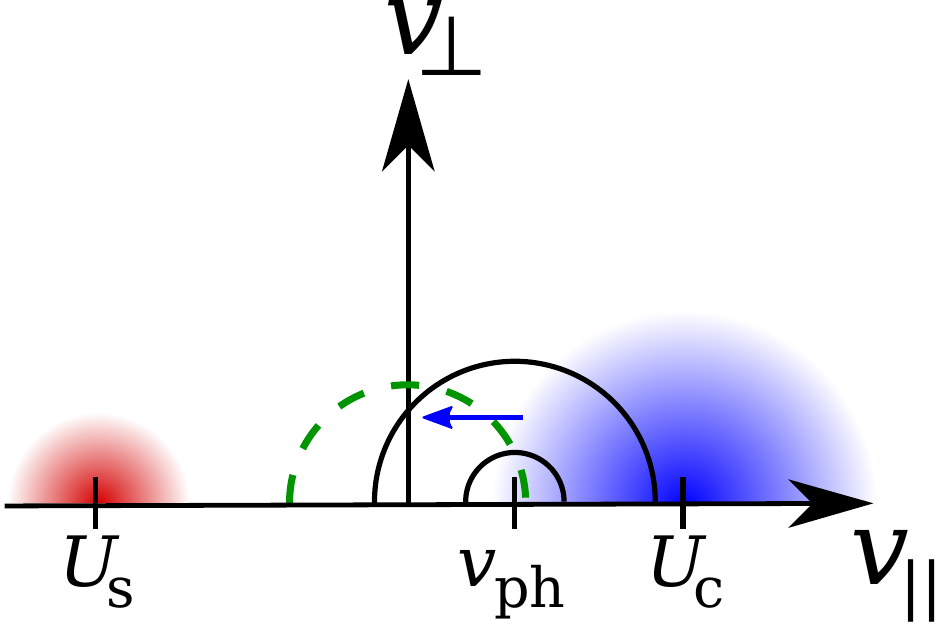}
\end{center}
\caption{
Schematic illustration of the quasi-linear diffusion in the electron/ion-acoustic instability and in the kinetic Alfv\'en heat-flux instability. The blue shaded area represents the core electron population with bulk velocity $U_{\mathrm c}$, and the red shaded area represents the strahl population with bulk velocity $U_{\mathrm s}$. The black semi-circles represent Eq.~(\ref{circle}) for either wave type with parallel phase speed $v_{\mathrm{ph}}$. The diffusive flux of Landau-resonant particles is shown by the blue arrow. The green-dashed semi-circle indicates $v_{\perp}^2+v_{\parallel}^2=\text{constant}$. In our geometrical convention with $v_{\mathrm{ph}}>0$, $U_{\mathrm s}<0$ here, which indicates that the diffusive flux points in the $-\vec B_0$ direction. We note that $U_{\mathrm s}<0$ still corresponds to the anti-sunward direction if $\vec B_0$ points towards the Sun.
}\label{fig_schematic_iaw_kaw}
\end{figure}

The electron/ion-acoustic instability is an example of an electrostatic instability driven by an asymmetry in the electron distribution function. It is driven by the Landau-resonance of electrons with the ion-acoustic mode \citep{fried61}.  In a magnetised plasma, it has maximum growth when $k_{\perp}=0$.

The electron/ion-acoustic instability can be driven if there is a non-zero current parallel to $\vec B_0$ in a plasma with a single proton and a single electron component. While the ion-acoustic wave is strongly Landau-damped in plasmas with $T_{\mathrm e}\approx T_{\mathrm p}$, a sufficiently large relative drift between the protons and the electrons leads to an instability. The dispersion relation of the ion-acoustic wave is given by \citep{gary93,verscharen17}
\begin{equation}\label{iaw_dr}
\omega_{k\mathrm r}\approx k_{\parallel}c_{\mathrm s}=k_{\parallel}\sqrt{\frac{3k_{\mathrm B}T_{\parallel\mathrm p}+k_{\mathrm B}T_{\parallel \mathrm e}}{m_{\mathrm p}}},
\end{equation}
where $c_{\mathrm s}$ is the ion-acoustic speed. The protons provide the wave inertia, while the proton and electron pressures provide the restoring force. In their wave evolution, protons behave like a one-dimensional adiabatic fluid, while the electrons are isothermal according to Eq.~(\ref{iaw_dr}). When the instability is only weakly unstable (i.e., small $\gamma_{\mathrm m}>0$), it is a long-wavelength electrostatic instability with $k\ll 1/\lambda_{\mathrm p}$. With increasing $\gamma_{\mathrm m}$, the unstable wave-vector space increases to $0<k\lesssim 1/\lambda_{\mathrm p}$ \citep{gary93}.

The quasi-linear evolution of $f_{0\mathrm e}$ under the action of the electron/ion-acoustic instability corresponds to the case shown in Fig.~\ref{fig_schematic_iaw_kaw}. If it is driven by the current between a single proton and a single electron component, the red strahl population in Fig.~\ref{fig_schematic_iaw_kaw} can be ignored. Due to the relative drift between the core population and the protons, Landau-resonant electrons diffuse towards smaller $v_{\parallel}$, leading to a reduction in the kinetic energy of the resonant electrons. This process drives the resonant ion-acoustic waves unstable at the expense of the relative drift between protons and electrons.

For the case of a single drifting electron component, \citet{gary93} gives an expression for the growth rate of the electron/ion-acoustic instability in a plasma with Maxwellian distribution functions under the assumption that $T_{\mathrm e}\gg T_{\mathrm p}$:
\begin{equation}\label{gamma_iaw}
    \gamma_k=\frac{\sqrt{\pi} \omega_{k\mathrm r}^3}{2|k_{\parallel}|^3}\left(\frac{m_{\mathrm p}}{k_BT_{\mathrm e}}\right)\frac{ k_{\parallel}  U_{\mathrm c}-\omega_{k\mathrm r}}{w_{\mathrm e}}\exp\left(-\frac{\left(\omega_{k\mathrm r}/k_{\parallel}-U_{\mathrm c}\right)^2}{w_{\mathrm e}^2}\right),
\end{equation}
where $U_{\mathrm c}$ is the bulk velocity of the single electron component in the proton reference frame.
Combining Eqs.~(\ref{iaw_dr}) and (\ref{gamma_iaw}) shows that $\gamma_k>0$ if $U_{\mathrm c}>v_{\mathrm {ph}}$, where $v_{\mathrm{ph}}\approx c_{\mathrm s}$. As shown in Fig.~\ref{fig_schematic_iaw_kaw}, the transition from $U_{\mathrm c}<v_{\mathrm {ph}}$ to $U_{\mathrm c}>v_{\mathrm {ph}}$ marks the transition from a diffusion that raises $v_{\parallel}$ (the blue arrow would be pointing to the right in this case) of the Landau-resonant electrons to a diffusion that lowers their $v_{\parallel}$ (blue arrow pointing to the left as shown). The growth rate according to Eq.~(\ref{gamma_iaw}) has a strong dependence on the electron temperature.

If the electron beam is very fast or the electrons are cold (i.e., $U_{\mathrm c}\gtrsim w_{\mathrm e}$), the dispersion relation of this unstable mode changes significantly from Eq.~(\ref{iaw_dr}). In this cold-plasma regime, the instability transitions into the classic Buneman electron/ion two-stream instability \citep{buneman59}. It corresponds to the $P=0$ mode in cold-plasma theory \citep{stix92} with $\omega_{k\mathrm r}\sim \omega_{\mathrm{pe}}$ and maximum growth at $k_{\parallel}\simeq \omega_{\mathrm{pe}}/U_{\mathrm c}$ when $U_{\mathrm c}\gg w_{\mathrm e}$.

In the solar wind, the persistent occurrence of sufficiently strong field-aligned currents to drive the electron/ion-acoustic instability via this mechanism is unlikely. For reference, the most intense current densities in the solar wind at 1\,au are typically $\sim 5\, \mathrm{nA/m}^{2}$ \citep{Podesta17:jgr}, and the corresponding net drift between ions and electrons is very small \citep{Vasko22:apjl}. However, a two-component electron configuration as shown in Fig.~\ref{fig_schematic_iaw_kaw} consisting of a core and strahl population enables the same instability mechanism. In an electron--proton plasma with core and strahl populations, the system is free from parallel currents if 
\begin{equation}\label{qncurrent}
n_{\mathrm s}U_{\mathrm s}+n_{\mathrm c}U_{\mathrm c}=n_{\mathrm p}U_{\mathrm p},
\end{equation} 
which is typically the case in the solar wind \citep{feldman75, salem22}.
In this configuration, the Landau-resonant interaction between unstable core electrons and ion-acoustic waves leads to a reduction of $v_{\parallel}$ of the resonant electrons. As this corresponds to a reduction in $U_{\mathrm c}$, the current-balance requirement from quasi-neutrality then also leads to a reduction in $|U_{\mathrm s}|$ \citep[see also][]{schroeder21}. Therefore, the electron/ion-acoustic instability is a candidate for the (indirect) regulation of the strahl heat flux in the solar wind \citep{gary78}. 
As shown in Fig.~\ref{fig_schematic_iaw_kaw}, this instability does not scatter strahl electrons into the halo, although such a behaviour is found in the solar wind \citep{stverak09}.

Ion-acoustic waves have been observed in the solar wind \citep{gurnett77,kurth79,gurnett91,pisa21}.  They often occur in sporadic bursts and at times when $T_{\parallel\mathrm e}>T_{\parallel \mathrm p}$ \citep{mozer21}. 
Near the Sun, the condition that $T_{\parallel\mathrm e}\gg T_{\parallel \mathrm p}$ can be satisfied in low-speed solar wind. At small heliocentric distances, the proton temperature remains correlated with the wind speed, but the electron temperature is anti-correlated with the wind speed, most likely due to the initial conditions in the corona \citep{halekas20, maksimovic_anticorrelation_2020}. The resulting conditions in slow-speed near-Sun solar wind thus favour the growth of the ion-acoustic wave. Indeed, Parker Solar Probe observes ion-acoustic waves under these conditions \citep{Mozer_2022}. The loose correlation between ion-acoustic waves with periods of enhanced electron temperatures suggests  that these waves may heat the core electrons. However, since high electron temperature itself favours the growth of the waves, the causality remains unclear. 

Near 1\,au, other instabilities often have lower thresholds than the electron/ion-acoustic instability though \citep{gary78,lemons79}. A direct stability analysis of measured electron distributions from Helios identifies a case that is unstable against the electron/ion-acoustic instability \citep{dum80}.  Strong ion-acoustic wave bursts are also found near magnetic switchbacks \citep{mozer21a}. The exact generation mechanism of these waves is unclear, as an ion/ion-acoustic instability is also a possible candidate for the generation of these waves \citep{mozer20,graham21}.

The current-driven electron/ion-acoustic instability is studied numerically in the context of laser-heated laboratory plasmas \citep{detering05}.

\subsection{Kinetic Alfvén heat-flux instability}

The kinetic Alfv\'en heat-flux instability is driven by the same instability mechanism as the electron/ion-acoustic instability shown in Fig.~\ref{fig_schematic_iaw_kaw}. Also in this instability, Landau-resonant core electrons diffuse towards smaller $v_{\parallel}$. The unstable wave mode in this case is the highly-oblique kinetic Alfv\'en wave, which is an electromagnetic plasma mode. It corresponds to the small-wavelength extension of the Alfv\'en wave in highly oblique propagation (i.e., $k_{\perp}\rho_{\mathrm p}\gtrsim 1$ and $k_{\perp}\gg k_{\parallel}$). Its dispersion relation is given by  \citep{howes06}
\begin{equation}
    \omega_{k\mathrm r}\approx \frac{k_{\parallel}v_{\mathrm A}k_{\perp}\rho_{\mathrm p}}{\sqrt{\beta_{\mathrm p}+\displaystyle \frac{2}{ 1+T_{\mathrm e}/T_{\mathrm p}}}}.
\end{equation}
In a Maxwellian electron--proton plasma, kinetic Alfv\'en waves undergo electron-Landau damping. With the introduction of a core--strahl configuration with sufficiently large $U_{\mathrm c}>0$, the plasma can achieve  $\partial f_{0\mathrm e}/\partial v_{\parallel}>0$ at $v_{\parallel}=v_{\mathrm{ph}}$, in which case the kinetic Alfv\'en wave is driven unstable. The quasi-neutrality condition in Eq.~(\ref{qncurrent}) enforces a simultaneous reduction of $|U_{\mathrm s}|$ when the instability reduces $U_{\mathrm c}$ like in the case of the electron/ion-acoustic instability. The kinetic Alfv\'en heat-flux instability has maximum growth at $k_{\perp}\lesssim 1/\rho_{\mathrm p}$ \citep{gary75a}.

Under typical solar-wind conditions with $\beta_j\sim 1$, the kinetic Alfv\'en  heat-flux instability has a significantly greater threshold than the parallel whistler heat-flux instability \citep[Section~\ref{sec_whistler_hf};][]{gary75a,gary75b}. In addition, this instability does not explain the observed scattering of strahl electrons into the halo population \citep{stverak09} since only core electrons diffuse in velocity space towards smaller $v_{\parallel}$ \citep{verscharen19}.

Kinetic Alfv\'en waves are often observed in the solar wind \citep{leamon98,bale05,chen10,salem12,safrankova19,roberts20}. However, their presence is generally neither attributed to ion-driven nor to electron-driven instabilities. Instead, they are interpreted as the small-wavelength extension of the Alfv\'enic cascade of solar-wind turbulence \citep{howes06,schekochihin09}.

The kinetic Alfv\'en heat-flux instability has  not been investigated extensively via numerical simulations. In contrast, kinetic Alfv\'en waves more generally have been the subject of  intense study. For example,  \citet{gary2004kinetic} compare linear theory and particle-in-cell simulations (albeit employing a low proton-to-electron mass ratio) of kinetic Alfv\'en waves to quantify the associated electron heating. Particle-in-cell and gyrokinetic simulations of kinetic Alfv\'en turbulence in the solar wind are used to investigate ion and electron heating \citep{howes08,howes11,parashar2015turbulent,hughes2017kinetic,groselj18,cerri19}.  Furthermore, kinetic Alfv\'en waves are routinely generated in particle-in-cell simulations of magnetic reconnection in conjunction with Hall physics in the diffusion region \citep{rogers2001role, shay2011super}.

\subsection{Langmuir instability and electron-beam instability}\label{sec_elstat}

\begin{figure}[ht]
\begin{center}
\includegraphics[width=0.6\textwidth]{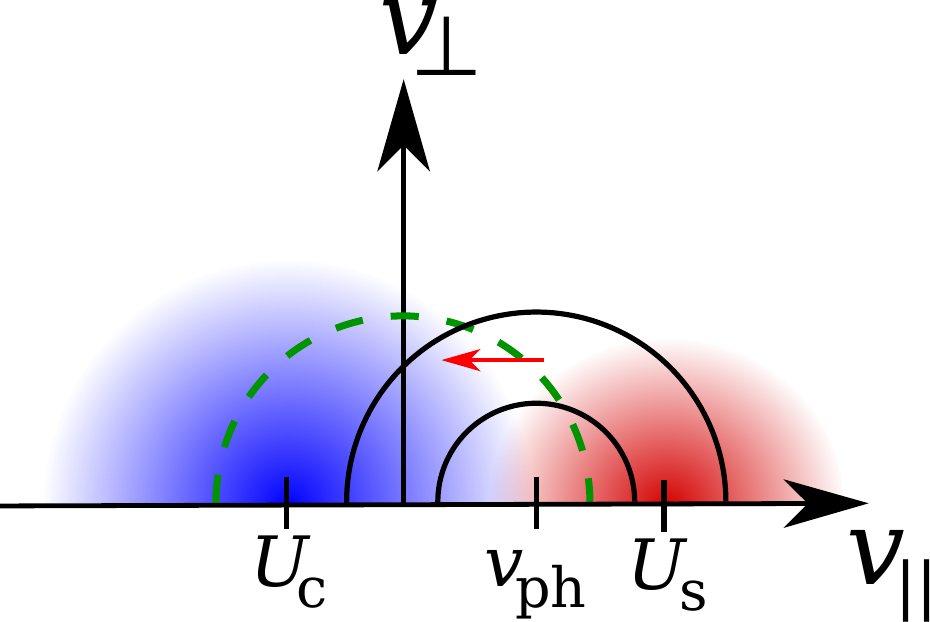}
\end{center}
\caption{
Schematic illustration of the quasi-linear diffusion in the Langmuir instability and  the electron-beam instability. The blue shaded area represents the core electron population with bulk velocity $U_{\mathrm c}$, and the red shaded area represents the strahl population with bulk velocity $U_{\mathrm s}$. The black semi-circles represent Eq.~(\ref{circle}) for either wave type with parallel phase speed $v_{\mathrm{ph}}$. The diffusive flux of Landau-resonant particles is shown by the red arrow.  The green-dashed semi-circle indicates $v_{\perp}^2+v_{\parallel}^2=\text{constant}$. 
}\label{fig_schematic_elstat}
\end{figure}

The Langmuir instability and the electron-beam instability are two examples of beam-driven electrostatic instabilities \citep[also called ``high-frequency electron/electron instabilities'';][]{gary85}. They are both driven by  Landau-resonant electrons, which requires that $\partial f_{0\mathrm e}/\partial v_{\parallel}>0$ at $v_{\parallel}=v_{\mathrm{ph}}$ when $v_{\mathrm{ph}}>0$. This configuration corresponds to a bump-on-tail distribution. Both instabilities have maximum growth when $k_{\perp}=0$.

In the relevant high-frequency range and assuming only a small modification to the real-part of the dispersion relation from any electron beam components, there are two solutions to the dispersion relation that become unstable. The Langmuir wave follows the dispersion relation
\begin{equation}
\omega_{k\mathrm r}\approx \sqrt{\omega_{\mathrm {pe}}^2+\frac{3}{2}k_{\parallel}^2w_{\mathrm c}^2}.
\end{equation}
If an electron-beam (strahl) component with bulk velocity $U_{\mathrm s}$ is present, the plasma also supports an electron-beam mode with \citep{gary78}
\begin{equation}\label{elbeamDR}
    \omega_{k\mathrm r}\approx k_{\parallel} U_{\mathrm s}.
\end{equation}

The criterion for distinguishing which of the two modes becomes unstable when $\partial f_{0\mathrm e}/\partial v_{\parallel} > 0 $ at $v_{\parallel} = v_{\mathrm {ph}}$ depends on the speed, temperature, and relative density of the beam \citep{oneil68}. Under unstable conditions, the Langmuir wave is the relevant mode if \citep{gary93} 
\begin{equation}
    \left(\frac{2U_{\mathrm s}}{w_{\mathrm s}}\right)^3\left(\frac{n_{\mathrm s}}{n_{\mathrm c}}\right)<1,
\end{equation}
and the electron-beam mode is the relevant mode if
\begin{equation}
    \left(\frac{2U_{\mathrm s}}{w_{\mathrm s}}\right)^3\left(\frac{n_{\mathrm s}}{n_{\mathrm c}}\right)\gtrsim 1.
\end{equation}
Both the Langmuir and the electron-beam instability have high frequencies (compared to $\Omega_{\mathrm p}$) and  wave numbers $k_{\parallel} \ll 1/\lambda_{\mathrm e}$, often even $k_{\parallel} \lesssim 1/\lambda_{\mathrm p}$, at maximum growth \citep{gary93}.

The quasi-linear evolution of $f_{0\mathrm e}$ under the action of the Langmuir instability or of the electron-beam instability is shown in Fig.~\ref{fig_schematic_elstat}. Landau-resonant strahl electrons diffuse towards smaller $v_{\parallel}$ in this configuration, while the core bulk velocity increases in order to guarantee current balance according to Eq.~(\ref{qncurrent}). This process leads to a flattening of the distribution function around $v_{\parallel}= v_{\mathrm{ph}}$. If the dispersion relation in Eq.~(\ref{elbeamDR}) were fulfilled exactly and the strahl distribution were symmetric around $v_{\parallel}=U_{\mathrm s}$, then $\partial f_{0\mathrm e}/\partial v_{\parallel}=0$ for Landau-resonant electrons at $v_{\parallel}=v_{\mathrm{ph}}$. In this case, the instability would not act. This illustrates the importance of subtle modifications to the dispersion relation due to the beam component itself so that $U_{\mathrm s}$ is slightly greater than $v_{\mathrm{ph}}$ in order to create an unstable configuration. 

Under certain parameter combinations, especially at large beam speeds, the Langmuir mode and the electron-beam mode couple in their dispersion relation \citep{gary93}.
Since the electron-beam mode's phase speed is approximately equal to the parallel speed of the resonant electrons, it typically grows over a wide range of frequencies, which is important in the Earth's foreshock, where bump-on-tail configurations occur frequently \citep{fuselier85}. However, this behaviour changes when the modes couple because then the dispersion relation becomes more complex.

In the cold-plasma limit (i.e., for very fast and cold beams with $U_{\mathrm s}\gg w_{\mathrm s}$ and $U_{\mathrm s}\gg w_{\mathrm c}$), the electron-beam instability corresponds to the classical electron two-stream instability. In this limit, the instability has $\omega_{k\mathrm r}\simeq \omega_{\mathrm{pe}}$ and a maximum growth rate of \citep{gary93}
\begin{equation}
    \gamma_{\mathrm m}\simeq \frac{\sqrt{3}}{2}\left(\frac{n_{\mathrm s}}{2n_{\mathrm  c}}\right)^{1/3}\omega_{\mathrm{pc}}.
\end{equation}

In low-$\beta_{\mathrm c}$ conditions, the Langmuir and electron-beam instabilities can have lower thresholds than other beam-driven instabilities. However, they both require a bump-on-tail configuration in order to be driven. The electron strahl does not generally generate such a non-monotonic $v_{\parallel}$-dependence of $f_{0\mathrm e}$. However, observations in the Earth's foreshock find evidence for the Langmuir instability when tenuous and fast electron beams are present, and for the electron-beam instability when denser and slower electron beams are present \citep{etcheto84,lacombe85,onsager90}.

Langmuir waves are often observed in the solar wind  at different heliospheric distances \citep{Kennel1980}. They frequently occur at the same time as narrow-band electromagnetic waves identified as whistler waves, potentially suggesting a common origin \citep{jagarlamudi_whistler_2021}. 

Two-dimensional, electromagnetic particle-in-cell simulations of a core-strahl electron configuration reveal that the electrostatic electron-beam instability also develops fluctuations in the perpendicular electric-field component, which scatter strahl electrons towards greater $v_{\perp}$ \citep{gary07}. A Fokker--Planck model of wave--particle interactions between an electron beam and the Langmuir instability suggests a similar process \citep{pavan13}.

\subsection{Electron/ion-cyclotron instability}

In low-$\beta_{j}$ plasmas, highly oblique electrostatic ion-cyclotron waves exist \citep{angelo62,stix92}. These modes occur in bands between the harmonics of the proton gyrofrequency \citep{gary93}: 
\begin{equation}
    m\Omega_{\mathrm p}<\omega_{k\mathrm r}<(m+1)\Omega_{\mathrm p},
\end{equation} 
where $m\ge 1$ is the integer harmonic order of the electrostatic ion-cyclotron wave. 

In a plasma consisting of a single electron and a single proton population, electrostatic ion-cyclotron waves can become unstable if there is a sufficiently large current given by a difference in the bulk speeds of the electrons and the protons parallel to $\vec B_0$ \citep{drummond62, kindel71}. Strong Landau-resonant interactions between the electrons and the harmonics of the ion-cyclotron wave are responsible for the driving of this instability.  The wave number at maximum growth typically fulfills $k_{\perp}\sim 1/\rho_{\mathrm p}$.

The quasi-linear evolution of $f_{0\mathrm e}$ is similar to the process described in Fig.~\ref{fig_schematic_elstat}, but where the strahl population is the only electron population. Since electrostatic ion-cyclotron waves are highly dispersive, $v_{\mathrm{ph}}$ depends strongly on $v_{\parallel}$ of the resonant electrons. 

The ion-cyclotron wave with $m=1$ has the lowest threshold. With increasing $U_{\mathrm c}$, harmonics with higher $m$ become unstable as well. Cyclotron-resonant interactions with protons compete with the Landau-resonant electron driving. Therefore, the instability threshold depends strongly on $T_{\mathrm p}$ \citep{gary93}. For $T_{\mathrm e}/T_{\mathrm p}\lesssim 10$, the electron/ion-cyclotron instability has a lower threshold than the current-driven electron/ion-acoustic instability. At $T_{\mathrm e}/T_{\mathrm p}\gtrsim 50$, the unstable ion-cyclotron branch of the dispersion relation merges with the unstable ion-acoustic branch, so that this instability loses its identity. 

As in the case of other current-driven instabilities, we expect that the introduction of a core-strahl configuration can also create the conditions necessary for the electron/ion-cyclotron instability in a plasma fulfilling Eq.~(\ref{qncurrent}). In this case, the instability mechanism requires $\partial f_{0\mathrm e}/\partial v_{\parallel}>0$ at $v_{\parallel}=v_{\mathrm {ph}}$ and would be the same as shown in Fig.~\ref{fig_schematic_elstat}. However, we are not aware of detailed studies of the conditions necessary for a core-strahl electron system to drive electrostatic ion-cyclotron waves unstable.

In order to overcome cyclotron-resonant proton damping, this instability is only relevant in plasma environments with  $\beta_{\parallel \mathrm p}\ll 1$. Therefore, the electron/ion-cyclotron instability is mostly  thought to occur within  low-$\beta_j$ environments such as the auroral ionosphere \citep{ashour78,bergmann84}. Driven by electrons and damped by protons, the unstable ion-cyclotron waves lead to efficient proton heating in this environment  \citep{okuda83,ashour84}.

A Fokker--Planck model of the current-driven electron/ion-cyclotron instability shows the presence of quasi-linear cyclotron-resonant diffusion effects on the proton distribution function, leading to its flattening in the resonance region \citep{Harvey1975}. These simulation results also confirm that the conditions for this instability to be excited are not commonly satisfied in typical solar-wind conditions.

\subsection{Electron/electron-acoustic instability}

\begin{figure}[ht]
\begin{center}
\includegraphics[width=0.6\textwidth]{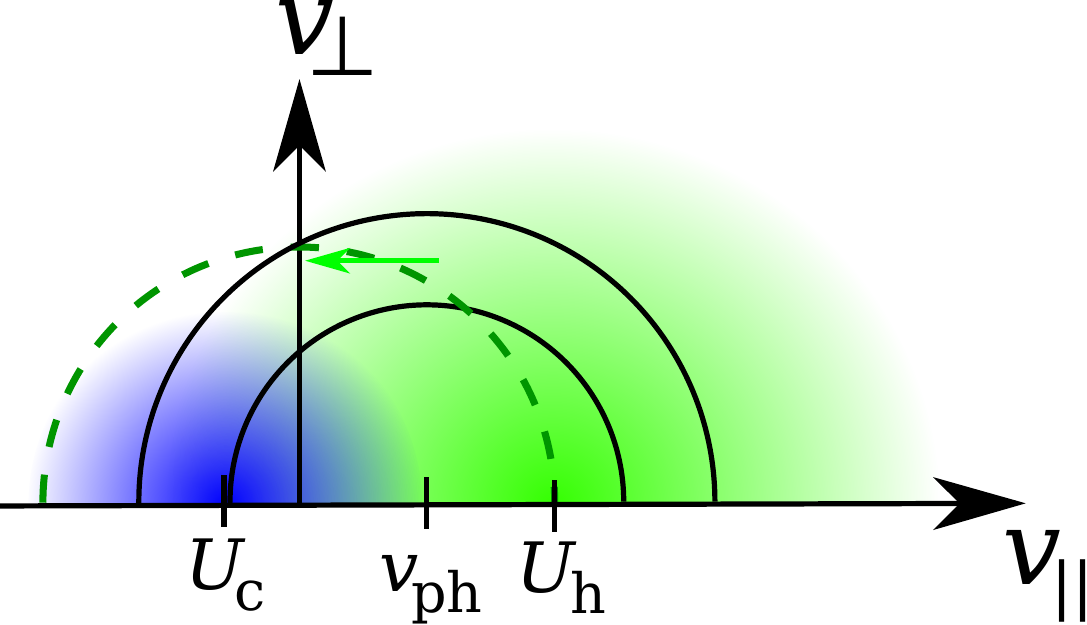}
\end{center}
\caption{
Schematic illustration of the quasi-linear diffusion in the electron/electron-acoustic  instability. The blue shaded area represents the core electron population with bulk velocity $U_{\mathrm c}$, and the green shaded area represents the halo population with bulk velocity $U_{\mathrm s}$. The electron/electron-acoustic instability requires $T_{\mathrm h}\gg T_{\mathrm c}$. The black semi-circles represent Eq.~(\ref{circle}) for electron-acoustic waves  with parallel phase speed $v_{\mathrm{ph}}$. The diffusive flux of Landau-resonant particles is shown by the green arrow.  The green-dashed semi-circle indicates $v_{\perp}^2+v_{\parallel}^2=\text{constant}$. 
}\label{fig_schematic_eaw}
\end{figure}

In a plasma consisting of protons and two electron populations, an additional electrostatic mode with properties similar to the ion-acoustic wave from Section~\ref{sec_iaw} emerges \citep{watanabe77}. This mode is called the electron-acoustic wave. In order for this mode to have a small damping rate, it is required that the two electron components have comparable densities but that one of the electron components is much hotter than the other  \citep{gary87}. We identify the hotter component with a possible halo population in the solar wind. 

If $n_{\mathrm h}\sim n_{\mathrm c}$, $T_{\mathrm h}\gg T_{\mathrm c}$, and $U_{\mathrm h}=U_{\mathrm c}=0$, the electron-acoustic wave has the dispersion relation \citep{watanabe77,gary87}
\begin{equation}\label{elawDR}
    \omega_{k\mathrm r}\approx \omega_{\mathrm{p c}}\sqrt{\frac{1+3k_{\parallel}^2\lambda_{\mathrm c}^2}{1+1/k_{\parallel}^2\lambda_{\mathrm h}^2}}.
\end{equation}
The cold electron component provides the wave inertia, while the high mobility of the hot electrons provides the restoring force of the electron-acoustic wave. At long wavelengths, $v_{\mathrm{ph}}$ is approximately proportional to $\sqrt{T_{\mathrm h}}$.

With the introduction of a sufficient relative drift speed between the core and halo populations, the electron-acoustic mode becomes unstable through the Landau-resonant interaction between halo electrons and the electron-acoustic mode. In this case, the mode still fulfills Eq.~(\ref{elawDR}), but in the frame of the core electrons, which provide the wave inertia. The wave number of the electron/electron-acoustic instability at maximum growth typically fulfills $1/\lambda_{\mathrm h}<k_{\parallel}<1/\lambda_{\mathrm c}$. It has maximum growth when $k_{\perp}=0$.

If the electron-acoustic mode is moderately unstable (i.e., $|\gamma_{\mathrm m}|\ll \omega_{k\mathrm r}$), its growth rate is given by \citep{gary93}
\begin{equation}\label{gammaEAW}
    \gamma_k\simeq \frac{\omega_{\mathrm{pc}}}{2|k_{\parallel}|^3\lambda_{\mathrm h}^2}\sqrt{\pi}\left(\frac{k_{\parallel}U_{\mathrm h}^{\prime}-\omega_{k\mathrm r}^{\prime}}{w_{\mathrm h}}\right)\exp \left(-\frac{\left(U_{\mathrm h}^{\prime}-\omega_{k\mathrm r}^{\prime}/k_{\parallel}\right)^2}{w_{\mathrm h}^2}\right)
\end{equation}
where $U_{\mathrm h}^{\prime}$ and $\omega_{k\mathrm r}^{\prime}$ are the bulk velocity of the halo population  and the wave frequency in the reference frame of the core population: $U_{\mathrm h}^{\prime}=U_{\mathrm h}-U_{\mathrm c}$ and $\omega_{k\mathrm r}^{\prime}=\omega_{k\mathrm r}-k_{\parallel}U_{\mathrm c}$. Eq.~(\ref{gammaEAW}) shows that, as for all Landau-resonant instabilities, the electron/electron-acoustic instability requires $U_{\mathrm h}>v_{\mathrm{ph}}$ when $v_{\mathrm{ph}}>0$ so that $\partial f_{0\mathrm e}/\partial v_{\parallel}>0$ at the value of $v_{\parallel}$ of the resonant electrons.

The quasi-linear evolution of $f_{0\mathrm e}$ under the action of the electron/electron-acoustic instability is shown in Fig.~\ref{fig_schematic_eaw}. Landau-resonant halo electrons diffuse towards smaller $v_{\parallel}$ in this configuration. 

The electron/ion-acoustic instability from Section~\ref{sec_iaw} has a lower threshold than the electron/electron-acoustic instability  unless  $T_{\mathrm h}\gg T_{\mathrm c}$ \citep{gary87}. In addition, a substantial $n_{\mathrm h}\sim n_{\mathrm c}$ is required. Both conditions are not generally fulfilled in the solar wind. However, cusp hiss fluctuations in the magnetospheric context are attributed to electron-acoustic fluctuations \citep{thomsen83,marsch85}. Electron-acoustic waves, typically with nonlinearly steepened wave forms, are reported in the Earth's inner magnetosphere, where the density of hot electrons can be greater than the density of cold electrons during hot plasma injections from the magnetotail \citep{Vasko17:grl,Dillard18:phpl}. In the near-Sun environment, the electron/electron-acoustic instability may be relevant for the regulation of the solar-wind heat flux \citep{sun21}.

Fully electromagnetic particle-in-cell simulations of the electron/electron-acoustic instability show that heating of the cold core component quenches the instability due to a breakdown of its requirement that $T_{\mathrm h}\gg T_{\mathrm c}$ \citep{lin85}. This appears to be the dominant nonlinear saturation mechanism for the electron/electron-acoustic instability \citep{gary93}.

\subsection{Whistler heat-flux instability}\label{sec_whistler_hf}

\begin{figure}[ht]
\begin{center}
\includegraphics[width=0.6\textwidth]{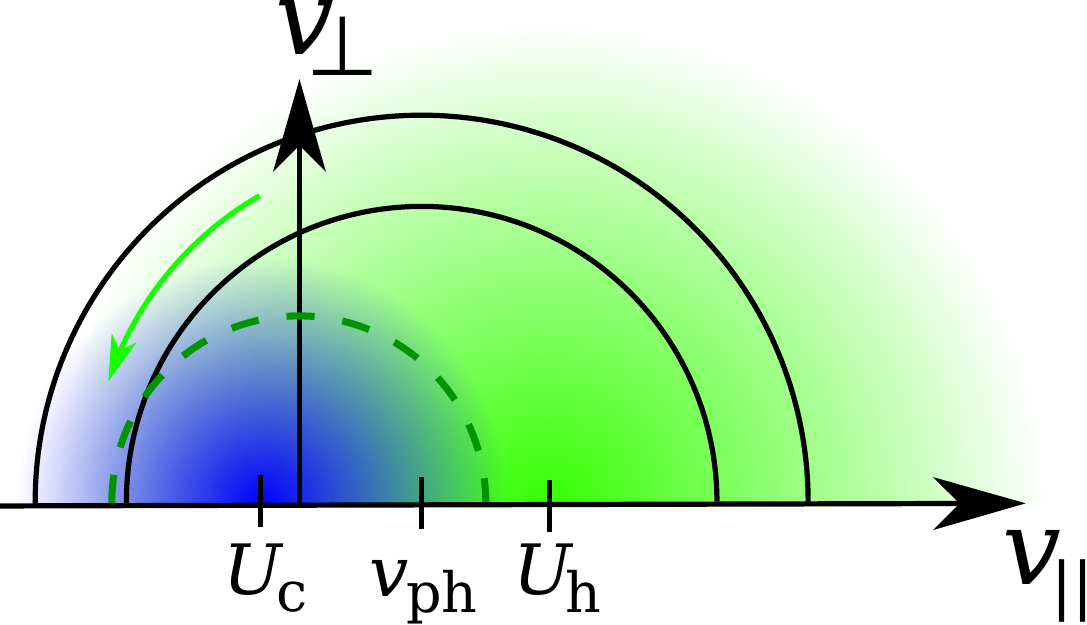}
\end{center}
\caption{
Schematic illustration of the quasi-linear diffusion in the whistler heat-flux  instability. The blue shaded area represents the core electron population with bulk velocity $U_{\mathrm c}$, and the green shaded area represents the halo population with bulk velocity $U_{\mathrm h}$. The black semi-circles represent Eq.~(\ref{circle}) for parallel-propagating fast-magnetosonic/whistler waves with parallel phase speed $v_{\mathrm{ph}}$. The diffusive flux of cyclotron-resonant particles is shown by the green arrow. The green-dashed semi-circle indicates $v_{\perp}^2+v_{\parallel}^2=\text{constant}$. 
}\label{fig_schematic_par_whistler}
\end{figure}

The whistler heat flux instability is a cyclotron-resonant instability of the electromagnetic fast-magnetosonic/whistler wave  \citep{gary75a,gary75b,schwartz80}. It has maximum growth at $k_\perp = 0$. Quasi-parallel fast-magnetosonic/whistler waves follow the approximate cold-plasma dispersion relation \citep{verscharen19a}
\begin{equation}\label{DRwhist}
    \frac{\omega_{k\mathrm r}}{\Omega_{\mathrm p}}\approx \frac{k_{\parallel}^2 d_{\mathrm{p}}^2}{2}\left(\sqrt{1+\frac{4\Omega_{\mathrm p}^2}{k_{\parallel}^2v_{\mathrm A}^2}}+1\right),
\end{equation}
when $\omega_{k\mathrm r}\ll |\Omega_{\mathrm e}|$.
The cyclotron-resonant electrons interact with the right-hand circularly polarised electric field of the fast-magnetosonic/whistler wave through the $n=-1$ resonance. According to Eq.~(\ref{rescond}), only electrons with $v_{\parallel}<0$ can fulfill the resonance condition when $v_{\mathrm{ph}}>0$.
In a plasma consisting of protons and two isotropic electron populations, the driving electron population must have a bulk velocity greater than $v_{\mathrm{ph}}$. Therefore, this instability is typically driven by a hot beam population like the halo. At maximum growth, $\Omega_{\mathrm p}\ll \omega_{k\mathrm r}\ll |\Omega_{\mathrm e}|$ and $1/d_{\mathrm p}\ll k_{\parallel}\lesssim 1/d_{\mathrm e}$.

Fig.~\ref{fig_schematic_par_whistler} shows the quasi-linear evolution of $f_{0\mathrm e}$ in the whistler heat-flux instability. In this case,  $0<v_{\mathrm{ph}}<U_{\mathrm h}$, and halo electrons with $v_{\parallel}<0$ resonate with the whistler wave. This setup guarantees that the flux of diffusing electrons in velocity space is directed as shown by the green arrow in Fig.~\ref{fig_schematic_par_whistler} as particles diffuse towards smaller values of $f_{0\mathrm e}$  \citep[see also][]{shaaban19}. The resonant electrons diffuse towards smaller $v_{\perp}$ but larger $|v_{\parallel}|$, while their $(v_{\perp}^2+v_{\parallel}^2)$ decreases. This decreasing kinetic energy is transferred into the resonant fast-magnetosonic/whistler waves. Increasing the halo density and temperature brings a larger number of electrons into resonance with the wave and thus leads to an increase in the growth rate. This instability is not a good candidate to explain the regulation of the strahl heat flux since the strahl does not provide a sufficient number of electrons at $v_{\parallel}<0$ when $U_{\mathrm s}>0$ \citep{verscharen19}.

With increasing halo speed, $\omega_{k \mathrm r}$ decreases compared to the traditional dispersion relation in Eq.~(\ref{DRwhist}). Figure 7 of \citet{gary85c}  compares the thresholds of the whistler heat-flux instability with the thresholds of the electron/ion-acoustic instability (see Section~\ref{sec_iaw}) and the electron-beam instability (see Section~\ref{sec_elstat}). Under typical solar-wind conditions, the whistler heat-flux instability has the lowest threshold of these instabilities. Only at large $n_{\mathrm h}/n_{\mathrm c}$, large $T_{\mathrm h}/T_{\mathrm c}$, and small $\beta_{j}$, the other instabilities can compete with the whistler heat-flux instability.

The instability mechanism of the whistler heat-flux instability is similar to the instability mechanism of the electron whistler anisotropy instability presented in Section~\ref{sec_whistler_ani}. The introduction of a halo anisotropy modifies the shape of $f_{0\mathrm e}$ in the velocity space occupied by resonant electrons. Consequently, the threshold of the whistler heat-flux instability decreases with increasing halo anisotropy $T_{\perp\mathrm h}/T_{\parallel \mathrm h}>0$ \citep[in general, the instability is sensitive to the shape of the halo distribution; see][]{abraham77,dum80}. 

The cyclotron-resonant halo driving competes with the cyclotron-resonant core damping of the fast-magnetosonic/whistler wave in this instability. Introducing a core anisotropy with $T_{\perp\mathrm c}/T_{\parallel \mathrm c}>0$ lowers the cyclotron-resonant core damping though and thus raises the growth rate.  Treatments of the whistler heat-flux instability in bi-Maxwellian and $\kappa$-distributed plasmas confirm this picture \citep{shaaban18,sarfraz20}.  Quasi-linear models of the whistler heat-flux and electron whistler anisotropy instability driven by a combination of heat flux and anisotropy are also available \citep{shaaban19b,vasko20}.

The thresholds of the whistler heat-flux instability have often been compared with the observed heat flux in the solar wind \citep{gary77,gary99b,tong19}. This instability is likely to operate near 1~au \citep{tong19}, but it appears unlikely to prove important near the Sun, where the halo is an almost negligible component of the distribution \citep{halekas20,halekas21,abraham22}. Conditions relevant for the driving of this instability also occur near interplanetary shocks and in the Earth's foreshock \citep{wilson09,wilson13,page21}. Quasi-parallel whistler waves are observed in solar-wind intervals with strong heat flux, supporting the links between core-halo heat flux and the whistler heat-flux instability \citep{lacombe14,stansby16,jagarlamudi20,tong19a}.
The contributions of core and halo anisotropies to the growth of the instability are confirmed observationally through the presence of a clear positive correlation between the occurrence of whistler waves and core/halo anisotropies \citep{jagarlamudi20}. Like the relative halo density, the core and halo anisotropies decrease with decreasing heliocentric distance, which is consistent with the observed lack of whistler waves at heliocentric distances $\lesssim 0.13\,\mathrm{au}$ \citep{cattell22}.

Particle-in-cell simulations of the whistler heat-flux instability reveal that beam scattering and core heating occur simultaneously due to the cyclotron-resonant wave--particle interactions \citep{lopez19}. The quasi-linear pitch-angle diffusion of the resonant electrons as shown in Fig.~\ref{fig_schematic_par_whistler} saturates quickly so that a significant heat-flux regulation is not expected. Numerical simulations confirm this expectation under typical solar-wind conditions \citep{kuzichev19}. Therefore, despite a significant amount of research into the action of the whistler heat-flux instability, its contribution to the observed heat-flux regulation in the solar wind is still not fully understood \citep[for observational constraints, see][]{feldman76a,feldman76b}.

\subsection{Oblique fast-magnetosonic/whistler instability}\label{sec_obl_whistler}

\begin{figure}[ht]
\begin{center}
\includegraphics[width=0.6\textwidth]{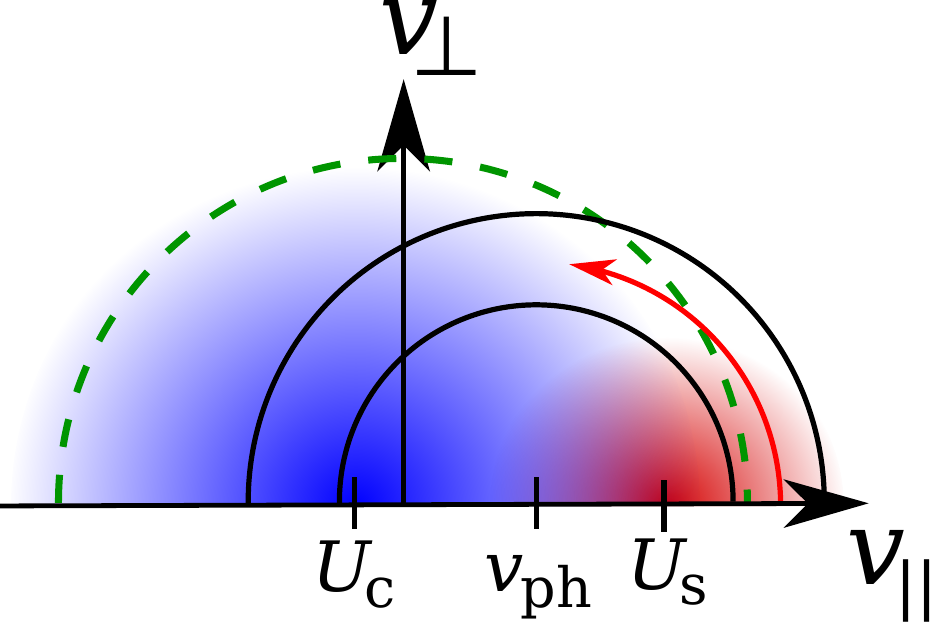}
\end{center}
\caption{
Schematic illustration of the quasi-linear diffusion in the oblique fast-magnetosonic/whistler instability and in the lower-hybrid fan instability. The blue shaded area represents the core electron population with bulk velocity $U_{\mathrm c}$, and the red shaded area represents the strahl population with bulk velocity $U_{\mathrm s}$. The black semi-circles represent Eq.~(\ref{circle}) for parallel-propagating fast-magnetosonic/whistler waves or lower-hybrid waves with parallel phase speed $v_{\mathrm{ph}}$. The diffusive flux of cyclotron-resonant particles is shown by the red arrow. The green-dashed semi-circle indicates $v_{\perp}^2+v_{\parallel}^2=\text{constant}$. 
}\label{fig_schematic_obl_whistler}
\end{figure}

When the fast-magnetosonic/whistler wave is not exactly parallel in propagation, it develops a non-zero left-circularly polarised component in its electric-field fluctuations. Unlike in the case of the whistler heat-flux instability presented in Section~\ref{sec_whistler_hf}, this allows electrons with $v_{\parallel}>0$ to resonate according to Eq.~(\ref{rescond}) with fast-magnetosonic/whistler waves with $v_{\mathrm {ph}} > 0$  via the $n=+1$ cyclotron resonance \citep{verscharen19}, provided $v_\parallel$ is sufficiently large. This enables strahl electrons with $v_{\parallel}>0$ to drive the oblique whistler wave unstable \citep{vasko19}. At maximum growth, $\Omega_{\mathrm p}\ll \omega_{k\mathrm r}\lesssim |\Omega_{\mathrm e}|$ and $1/d_{\mathrm p}\ll k_{\parallel}\lesssim 1/d_{\mathrm e}$.

The quasi-linear evolution of $f_{0\mathrm e}$ in the oblique fast-magnetosonic/whistler instability is shown in Fig.~\ref{fig_schematic_obl_whistler}. If the strahl distribution is isotropic, the resonant electrons diffuse towards larger $v_{\perp}$ and smaller $v_{\parallel}$ given that $0<v_{\mathrm{ph}}<U_{\mathrm s}$. In this case, $(v_{\perp}^2+v_{\parallel}^2)$ of the resonant strahl electrons decreases, corresponding to a loss of kinetic energy and thus a transfer of energy into the growing fast-magnetosonic/whistler waves. 

Due to its obliqueness, the fast-magnetosonic/whistler wave also possesses fluctuations in the electric-field component $E_{\parallel}$ parallel to $\vec B_0$. Therefore, the cyclotron-resonant driving by resonant strahl electrons competes not just with cyclotron-resonant damping by core electrons via $n=-1$ but also with Landau-resonant damping by core electrons via $n=0$. These competitions between driving and damping define two regimes of the oblique fast-magnetosonic/whistler instability: a high-$\beta_{\parallel\mathrm c}$ regime and a low-$\beta_{\parallel\mathrm c}$ regime. 

In the high-$\beta_{\parallel \mathrm c}$ regime (i.e., when $w_{\parallel \mathrm c}\gtrsim v_{\mathrm{Ae}}/2$), the competition between cyclotron-resonant strahl driving and Landau-resonant core damping determines the instability threshold. The oblique fast-magnetosonic/whistler wave is unstable in this case if \citep{verscharen19}
\begin{equation}\label{oblFMW1}
    U_{\mathrm s}\gtrsim \left[2\frac{n_{\mathrm c}}{n_{\mathrm s}}\sqrt{\frac{T_{\parallel \mathrm s}}{T_{\parallel \mathrm c}}}v_{\mathrm {Ae}}^2w_{\parallel\mathrm c}^2\frac{\left(1+\cos\theta\right)}{\left(1-\cos\theta\right)\cos\theta}\right]^{1/4},
\end{equation}
where $\theta$ is the angle between the wave vector and background magnetic field.

In the low-$\beta_{\parallel \mathrm c}$ regime (i.e., when $w_{\parallel \mathrm c}\lesssim v_{\mathrm{Ae}}/2$), the competition between cyclotron-resonant strahl driving,  cyclotron-resonant core damping, and Landau-resonant core damping determines the instability threshold. The oblique fast-magnetosonic/whistler wave is unstable in this case if
\begin{equation}\label{oblFMW2}
    U_{\mathrm s}\gtrsim 3w_{\parallel\mathrm c}.
\end{equation}
In the low-$\beta_{\parallel\mathrm c}$ regime, $\omega_{k\mathrm r}\approx |\Omega_{\mathrm e}|/2$,  $k_{\parallel}\approx |\Omega_{\mathrm e}|/2w_{\parallel\mathrm c}$,  and $\theta = 60^\circ$ at maximum growth.

Eqs.~(\ref{oblFMW1}) and (\ref{oblFMW2}) have been tested successfully against numerical solutions to the linear Vlasov--Maxwell dispersion relation for typical solar-wind parameters \citep{verscharen19}. 

Statistical comparisons of instability thresholds with electron measurements in the solar wind from Wind \citep{verscharen19} and from Parker Solar Probe \citep{halekas21} show that the strahl parameters are limited by Eq.~(\ref{oblFMW1}) to the stable parameter space.  However, a recent analysis of Parker Solar Probe and Helios data suggests that the strahl very rarely reaches the threshold in the inner heliosphere, so that the importance of this instability is now put into question \citep{jeong22b}. This finding is consistent with the observed lack of fast-magnetosonic/whistler waves in Parker Solar Probe data at  heliocentric distances $\lesssim 0.13\,\mathrm{au}$ \citep{cattell22}. Moreover, the majority of the fast-magnetosonic/whistler waves observed farther away from the Sun have a quasi-parallel direction of propagation with respect to the magnetic field  \citep{Kretzschmar2021}. Therefore,  other mechanisms than the self-induced scattering of strahl electrons by the oblique fast-magnetosonic/whistler instability may thus be needed to explain the observed scattering of strahl electrons into the halo population \citep[e.g., the interaction with pre-existing fast-magnetosonic/whistler waves;][]{vocks05,vocks09,Pierrard11,jagarlamudi_whistler_2021,cattell21,cattell21b}. Moreover, Bernstein and ion-acoustic waves become more dominant than fast-magnetosonic/whistler waves in the very inner heliosphere, suggesting a transition into an electrostatic regime which could affect the electron distributions near the Sun \citep[][]{malaspina21,mozer21}. The observed Bernstein and ion-acoustic waves occur in very specific types of solar wind: Bernstein waves occur predominantly  in quiet wind with magnetic field close to the ideal Parker spiral, while ion-acoustic waves occur predominantly in slow solar wind. This correlation underlines the importance of  the careful investigation of the electron distribution's evolution as a function of wind parameters.

Numerical evaluations of the quasi-linear diffusion equation confirm that the oblique fast-magnetosonic/whistler instability scatters resonant strahl electrons as shown in Fig.~\ref{fig_schematic_obl_whistler} \citep{jeong20,sun21}. Numerical particle-in-cell simulations also confirm this evolution under conditions consistent with observed solar-wind parameters  \citep{micera20}.

\subsection{Lower-hybrid fan instability}

At very large angles of propagation ($k_{\parallel}^2/k^2<m_{\mathrm e}/m_{\mathrm p}$), the fast-magnetosonic/whistler wave solution of the Vlasov--Maxwell dispersion relation is also known as the lower-hybrid mode. Its wave frequency asymptotes to the lower-hybrid frequency \citep{verdon09}
\begin{equation}
    \omega_{k\mathrm r}\approx \omega_{\mathrm{LH}}\equiv \frac{\omega_{\mathrm{pp}}}{\sqrt{1+\displaystyle \frac{\omega_{\mathrm{pe}}^2}{\Omega_{\mathrm e}^2}}}
\end{equation}
in the low-$\beta_{j}$ limit. Landau-resonant core damping strongly suppresses the lower-hybrid mode \citep{lakhina79}, so that it becomes most relevant when $\beta_{\mathrm c}\ll 1$, in which case the mode becomes largely electrostatic \citep{marsch83}.

In this low-$\beta_{\mathrm c}$ case, the lower-hybrid wave can be driven unstable by strahl electrons via the $n=+1$ cyclotron resonance like the oblique fast-magnetosonic/whistler instability \citep[see Section~\ref{sec_obl_whistler}; ][]{omelchenko94,krafft03,shevchenko10}. At maximum growth, the lower-hybrid fan instability has $\omega_{k\mathrm r}\approx \omega_{\mathrm{LH}}$ and $k_{\parallel}\approx \left(\omega_{\mathrm{LH}}+|\Omega_{\mathrm e}|\right)/U_{\mathrm s}$.

The instability mechanism for the lower-hybrid fan instability is the same as the mechanism driving the oblique fast-magnetosonic/whistler instability shown in Fig.~\ref{fig_schematic_obl_whistler}. Since the cyclotron-resonant diffusion leads to a fan-like widening of the strahl component in the perpendicular direction, this instability received the name lower-hybrid fan instability. 

Numerical Hamiltonian simulations of the lower-hybrid fan instability confirm the importance of both cyclotron-resonant and Landau-resonant interactions between electrons and waves \citep{krafft05,krafft06}. In its nonlinear stage, the lower-hybrid fan instability is prone to strong wave trapping that is not captured by quasi-linear theory.

\subsection{Electron-deficit whistler instability}

\begin{figure}[ht]
\begin{center}
\includegraphics[width=0.6\textwidth]{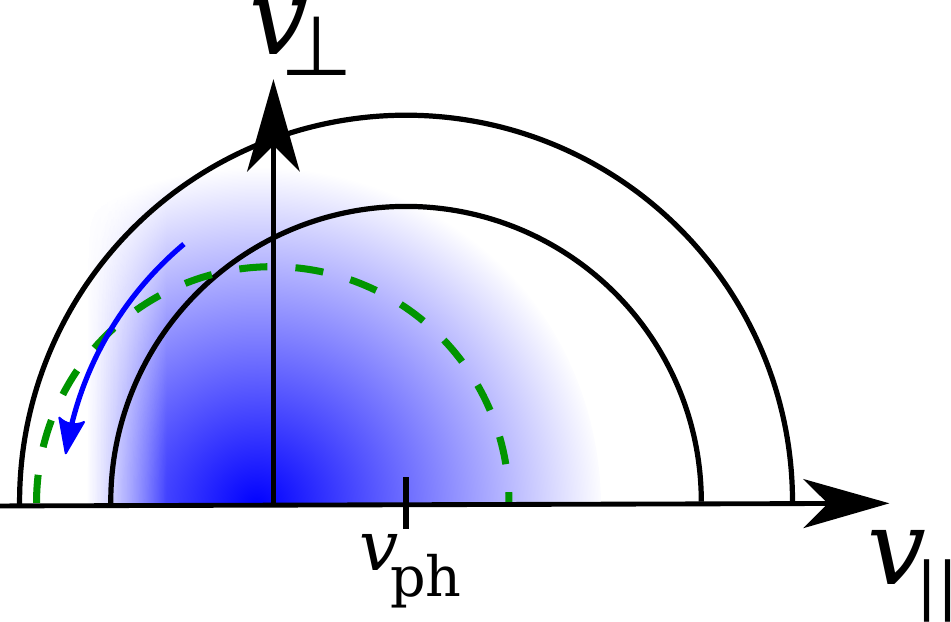}
\end{center}
\caption{
Schematic illustration of the quasi-linear diffusion in the electron-deficit whistler instability. The blue shaded area represents the electron population with a deficit at $v_{\parallel}<0$. The black semi-circles represent Eq.~(\ref{circle}) for parallel-propagating fast-magnetosonic/whistler waves with parallel phase speed $v_{\mathrm{ph}}$. The diffusive flux of cyclotron-resonant particles is shown by the blue arrow. The green-dashed semi-circle indicates $v_{\perp}^2+v_{\parallel}^2=\text{constant}$. 
}\label{fig_schematic_deficit}
\end{figure}

Up until this point, we have discussed the canonical examples of electron instabilities driven by anisotropy, drifts, or beams in the electron distribution function. However, other deviations from thermodynamic equilibrium are also able to drive instabilities if the deformation of the distribution is  sufficiently strong in the range of resonant velocities \citep{dum80}. 

One example of such a deformation of $f_{0\mathrm e}$ is the sunward deficit in the electron distribution \citep{halekas20,abraham22}. As expected in exospheric models of the solar wind \citep{lemaire71,Pierrard96,maksimovic01}, the interplanetary potential reflects electrons that leave the Sun with a kinetic energy below a cut-off value that depends on the potential. These reflected electrons return towards the Sun and form part of the sunward half of the electron distribution in the inner heliosphere. Electrons above the cut-off energy do not return. If collisions and other scattering mechanisms are neglected, a sharp cut-off is thus expected on the sunward side of the electron distribution, marking the separation between the reflected and the (missing) electrons that have escaped the potential. This cut-off has been observed in the form of a sunward deficit in the electron distribution function in data from Parker Solar Probe \citep{bercic21a,halekas21a}.

The sunward deficit can create conditions in which electrons near the cut-off  resonantly interact with parallel fast-magnetosonic/whistler waves such that they lose their kinetic energy and drive the wave unstable \citep{bercic21}. This interaction leads to the electron-deficit whistler instability. The instability at maximum growth has $\Omega_{\mathrm p}\ll \omega_{k\mathrm r} \ll |\Omega_{\mathrm e}|$ and for $k_{\perp}=0$. The wave number at maximum growth depends on the parallel velocity of the deficit in velocity space according to Eq.~(\ref{rescond}). The properties of this instability are still under study, but \citet{bercic21} suggest that the wave number at maximum growth is typically of order $\sim 1/d_{\mathrm e}$.

The quasi-linear evolution of $f_{0\mathrm e}$ in the electron-deficit whistler instability is shown in Fig.~\ref{fig_schematic_deficit}. The deficit is located at $v_{\parallel}<0$, where it modifies the relative alignment between the pitch-angle gradients of $f_{0\mathrm e}$ and the direction of the diffusive flux of resonant electrons  (locally tangent to the black semi-circles). Resonant electrons diffuse towards smaller $v_{\perp}$ and larger $|v_{\parallel}|$. The parallel fast-magnetosonic/whistler wave with phase speed $v_{\mathrm{ph}}$ grows at the expense of the energy of the diffusing electrons. This quasi-linear process fills up the electron deficit.

High-cadence and high-resolution measurements of the electron distribution function from Solar Orbiter show pronounced deficits at times when pronounced amplitudes of quasi-parallel fast-magnetosonic/whistler waves are seen \citep{bercic21}. This observation suggests the sporadic occurrence of the electron-deficit whistler instability in the solar wind. Since the sunward electron deficit is more pronounced near the Sun, a systematic study of the this instability would be worthwhile in data from Parker Solar Probe and Solar Orbiter at small heliocentric distances.

The observed deficit, rather than forming a sharp cutoff at a specific $v_{\parallel}$ as shown in Fig.~\ref{fig_schematic_deficit}, also extends to larger pitch angles, encompassing locally mirroring electrons with largely perpendicular velocities. This observation could indicate that the operation of the instability has already resulted in diffusion from perpendicular to parallel velocities, or it could indicate that those regions of phase space are also unpopulated as suggested by \citet{halekas21a}.  

Simulations of the electron-deficit whistler instability are not available yet.

\subsection{The impact of ion beams on electron-driven instabilities}

If the plasma consists of one electron population and multiple ion populations, relative drifts between the electrons and the ion populations can drive instabilities. The current balance in a system consisting of an electron component and two proton components demands
\begin{equation}
    n_{\mathrm e}U_{\mathrm e}=n_{\mathrm{pc}}U_{\mathrm{pc}}+n_{\mathrm{pb}}U_{\mathrm{pb}},
\end{equation}
where the subscript $\mathrm e$ refers to the single electron species, $\mathrm{pc}$ to the proton core, and $\mathrm{pb}$ to the proton beam. If such a proton beam--core configuration with $U_{\mathrm{pc}}\neq U_{\mathrm{pb}}$ exists, then $U_{\mathrm e}\neq U_{\mathrm{pc}}$; i.e., there is a non-zero drift between the proton core and the electrons. If the beam is sufficiently dense and fast in the proton-core frame, the fast-magnetosonic/whistler wave can be driven through cyclotron-resonant wave--particle interactions with the electrons \citep{akimoto87}. The mechanism is similar to the one shown in Fig.~\ref{fig_schematic_par_whistler} for the whistler heat-flux instability, where the population marked as the halo now corresponds to the only electron population. For isotropic electrons, this instability still requires that $U_{\mathrm e}>v_{\mathrm{ph}}$.  The separation between $U_{\mathrm e}$ and $v_{\mathrm{ph}}$ depends on the proton-beam  and proton-core properties.  The same mechanism is also potentially able to drive other instabilities presented in Section~\ref{Inst_asy}.

\section{Non-resonant instabilities}\label{Inst_nonres}

The instabilities discussed thus far are characterised by resonant wave--particle interactions that lead to quasi-linear diffusion of resonant electrons in velocity space. Another family of instabilities is characterised by non-resonant plasma processes. These non-resonant instabilities often also exist in fluid plasma models \citep{verscharen19a}. Our quasi-linear model does not apply to this family of instabilities. Therefore, we give a brief summary of two important examples only: the electron mirror-mode instability and the non-propagating firehose instability.

\subsection{Electron mirror-mode instability}

The mirror-mode instability is an example for a non-resonant, anisotropy-driven instability. It corresponds to the oblique non-propagating slow mode  with $\omega_{k\mathrm r}=0$ \citep{chandrasekhar58,barnes66,basu84,verscharen17}. Due to the polarisation of the non-propagating slow mode, the mirror-mode instability exhibits a significant component of magnetic-field fluctuations $\delta B_{\parallel}\equiv\delta \vec B \cdot \vec B_0/B_0$ parallel to $\vec B_0$. The fluctuations in $\delta B_{\parallel}$ are anti-correlated with the fluctuations in $\delta n_{\mathrm e}$. Trapping of slow ($v_{\parallel}\approx 0$) particles through the mirror force plays an important role in the nonlinear evolution of the mirror-mode instability \citep{southwood93}. However, particle trapping is not captured by our quasi-linear framework, and our requirement that  $\omega_{k\mathrm r}\gg \gamma_K$ is violated in the mirror-mode instability.

Unlike the resonant instabilities discussed in Sections~\ref{Inst_ani} and \ref{Inst_asy}, the mirror-mode instability is less sensitive to the shape of the distribution function in a defined narrow part of velocity space. Instead, its stability depends on the total pressure anisotropy of the system. This point is also illustrated by the analytical threshold for the mirror-mode instability which depends on the pressure contributions of both species in an electron--proton plasma: the mirror mode  is unstable if \citep[for a gyrokinetic derivation, see][]{verscharen19}
\begin{equation}\label{mirror1}
\beta_{\perp\mathrm p}\left(\frac{T_{\perp \mathrm p}}{T_{\parallel \mathrm p}}-1\right)+\beta_{\perp\mathrm e}\left(\frac{T_{\perp \mathrm e}}{T_{\parallel \mathrm e}}-1\right)>1.
\end{equation}
This type of analysis has also been extended to the case in which there are more than two particle species present \citep{hall79,hellinger07,chen16}.

Eq.~(\ref{mirror1}) illustrates that the mirror-mode instability can be driven unstable by both ions and electrons \citep[see also][]{migliuolo86}. The wave number at maximum growth shows an interesting transition between ion scales ($k\lesssim 1/d_{\mathrm p}$) and electron scales ($k\lesssim 1/d_{\mathrm e}$) depending on the species with the dominant anisotropy \citep{hellinger18}. 
The saturation of the mirror-mode instability happens via a fluid-level rearrangement of the plasma that reduces the overall pressure anisotropy \citep{kivelsen96,rincon15,riquelme15}.

Linear theory predicts that, for the same $\beta_{\parallel\mathrm e}$ and $T_{\perp\mathrm e}/T_{\parallel\mathrm e}$ in a bi-Maxwellian plasma, the  electron whistler anisotropy instability discussed in Section~\ref{sec_whistler_ani} generally has higher growth rates and lower thresholds than the oblique mirror-mode instability \citep{gary06}. However, two-dimensional particle-in-cell simulations show that both instabilities compete, and the oblique mirror-mode instability often becomes dominant in the non-linear phase after the parallel electron whistler anisotropy instability saturates \citep{hellinger18}. \citet{hellinger18} also note that the importance of the mirror-mode instability potentially increases in more realistic three-dimensional simulations due to the higher degrees of freedom in such a system.\footnote{Surprisingly, \citet{Sarfraz2021} perform two-dimensional simulations with the same physical parameters used by \citet{hellinger18} and do not observe any oblique modes. However, their simulation box is much smaller than that employed by \citet{hellinger18} and, as a consequence, the growth of the mirror instability may have been inhibited.}

\subsection{Non-propagating electron firehose instability}\label{oblFH}

The non-propagating firehose instability is another non-resonant instability driven by anisotropy. It corresponds to a non-propagating (i.e., $\omega_{k\mathrm r}=0$) solution of the oblique Alfv\'en-wave branch \citep{li00}. If the plasma-pressure anisotropy with $p_{\parallel}>p_{\perp}$ is sufficiently large, the magnetic tension is unable to provide a sufficient restoring force for the propagation of the Alfv\'en wave anymore, and the mode becomes aperiodic \citep{squire16}. In this context, we define the total pressures $p_{\perp}\equiv \sum_j n_jk_{\mathrm B}T_{\perp j}$ and $p_{\parallel}\equiv \sum_j n_jk_{\mathrm B}T_{\parallel j}$. Like in the case of the mirror-mode instability, the instability threshold of the non-propagating firehose instability depends on both electron and proton pressure contributions. In an electron--proton plasma, the non-propagating firehose mode is unstable if \citep[for a gyrokinetic derivation, see][]{verscharen19}
\begin{equation}
    \beta_{\parallel\mathrm p}-\beta_{\perp\mathrm p}+\beta_{\parallel\mathrm e}-\beta_{\perp\mathrm e}>2.
\end{equation}
If multiple species are present, this condition can be extended to
\citep{kunz15,chen16}
\begin{equation}
\sum \limits_j \left(\beta_{\parallel j}-\beta_{\perp j}+\frac{8\pi m_j n_j U_j^2}{B_0^2}\right)>2,
\end{equation}
which also accounts for the contributions of relative drifts to the total parallel pressure \citep{gary75a}.  Therefore, the non-propagating firehose instability can also be driven unstable in the presence of isotropic or anisotropic beam populations. At maximum growth, the non-propagating electron firehose instability has $k\sim 1/d_{\mathrm e}$.

The non-propagating firehose  has a significantly lower threshold than the propagating electron firehose instability under typical solar-wind conditions \citep{gary03,lazar22}.
The presence of suprathermal electron populations with properties consistent with observed solar-wind conditions lowers the threshold of the non-propagating firehose instability even further \citep{shaaban19a}.
In the expanding solar wind, conservation of the first adiabatic invariant (magnetic moment) in a decreasing magnetic field naturally increases $T_{\parallel\mathrm e}/T_{\perp\mathrm e}$  and drives the distribution toward the firehose instability thresholds \citep{Innocenti_2020}. If this process happens faster than Coulomb collisions can moderate the anisotropy, then the non-propagating firehose instability can be triggered.

The non-propagating electron firehose instability thresholds  constrain slow wind (but not fast wind) core electrons  to the stable parameter space in electron measurements from Helios, Cluster, and Ulysses \citep{stverak08}. Electrons measured by Parker Solar Probe during encounters 1 to 9 are stable and far from the thresholds of the non-propagating firehose instability \citep{cattell22}.

Particle-in-cell simulations of the non-propagating electron firehose instability show that nonlinear wave--wave interactions play an important role during the nonlinear stage of the instability \citep{camporeale08,hellinger14}. Highly oblique fluctuations grow initially, which then couple to modes with less oblique wave vectors. The interplay of the various modes leads to a situation in which the plasma ``bounces'' around the marginal stability threshold in parameter space. An anisotropy in suprathermal electron populations relaxes more quickly than an anisotropy in thermal electron populations during the instability's nonlinear evolution \citep{lopez19a}, which is also consistent with the observation of lower halo anisotropies \citep{stverak08}. Fully kinetic, expanding-box simulations demonstrate that the non-propagating electron firehose instability arises self-consistently in the expanding solar wind due to the conservation of the magnetic moment \citep{innocenti2019onset}. 
After onset, the firehose instability and its nonlinear evolution compete against the ongoing expansion to drive the system between stability and instability: also in expanding-box simulations, the electrons bounce around the marginal stability threshold in parameter space.

\section{Conclusions and open questions}\label{Conclusions}

We develop a semi-graphical framework for the analysis of resonant micro-instabilities based on the equations of quasi-linear theory. We apply this framework to electron-driven instabilities with relevance to the solar wind. With the help of this description, we discuss instabilities driven by temperature anisotropy and reflectional asymmetry in the electron distribution function.

Electrons make an important contribution to the overall dynamics and energetics of the solar wind through their pressure gradient and through their heat flux. Micro-instabilities modify the kinetic properties of the electron distribution function locally and thus have a local impact on the electron contributions to the global dynamics and energetics of the solar wind. Once triggered, they regulate the temperatures, temperature anisotropies, field-parallel currents, or heat flux. 
Throughout his career, Peter Gary has made groundbreaking contributions to the study of electron-driven instabilities in the solar wind, especially from the perspective of linear Vlasov--Maxwell theory and nonlinear plasma simulations. 

Measurements with modern space missions have revealed a number of open questions regarding the action and role of electron-driven instabilities in the solar wind. Here, we list a selection of these topics that we consider important for future research:
\begin{enumerate}
    \item While most linear-theory calculations assume a homogeneous and steady-state background plasma, the solar wind is far from homogeneity and a steady state. The plasma exhibits variations and inhomogeneities on a broad range of scales \citep{verscharen19a}. The action of  micro-instabilities in such a turbulent and variable plasma is not understood. Most relevant electron-driven instabilities  act on small electron scales ($\sim d_{\mathrm e}$ or $\sim \rho_{\mathrm e}$) at which the amplitude of the ubiquitous turbulent fluctuations in the solar wind is small. Therefore, they are likely to experience a less variable background on the relevant scales than ion-driven instabilities. Moreover, interactions between electrons and turbulent fluctuations also modify the electron distribution, so that a separation between the effects of micro-instabilities and the effects of turbulent dissipation is complex.
    \item The solar wind is an expanding plasma flow. Therefore, the background parameters of the system change as a parcel of solar-wind plasma travels through the heliosphere with its local bulk velocity. This global large-scale evolution modifies the kinetic structure of the particle distributions. Through numerical models such as expanding-box simulations \citep{innocenti2019semi,Innocenti_2020,micera21} or semi-analytical kinetic approaches \citep{sun21,jeong22a}, a simultaneous treatment of instabilities and expansion effects has become possible. In this context, the implementation of the regulating impact of electron-driven instabilities in global solar-wind models is an important goal for our understanding of the solar wind \citep[e.g.,][]{chandran11}. Despite progress in our numerical capabilities, a fully self-consistent treatment of the large-scale expansion and electron-driven instabilities still lies far in the future.
    \item Electron-driven instabilities are often treated in isolation. In reality, however, it is likely that electron-driven and ion-driven instabilities co-exist, depending on the mechanisms that create the driving deviations of the particle distributions from thermodynamic equilibrium. Numerical simulations that resolve both ion and electron processes and scales are a crucial tool for the understanding of the nonlinear stages of these combined instabilities \citep{schriver90,riquelme18,riquelme22}. For example, shearing particle-in-cell simulations show that unstable electron-scale fluctuations can grow inside the unstable ion-scale fluctuations in systems with driven anisotropy \citep{riquelme16}. Fully kinetic particle-in-cell simulations show that the electron firehose instabilities develop concurrently with the ion firehose instabilities \citep{lopez22}, which is a likely scenario since both species develop a temperature anisotropy due to solar-wind expansion. This effect increases the growth rate of the ion firehose instabilities compared to the linear prediction for the ion-driven firehose instabilities alone. We require such a combined description of electron-driven and ion-driven instabilities under realistic solar-wind conditions.
    \item Although some of the instabilities discussed in this review are able to regulate the heat flux of the electron distribution, recent research suggests that the solar-wind plasma rarely exceeds the linear instability thresholds for heat-flux driven instabilities \citep{horaites18,schroeder21,jeong22b}. These findings highlight the uncertainty in our understanding of the \emph{importance} of electron-driven instabilities in the solar wind. For all instabilities, it is crucial to investigate (a) how often they occur in the solar wind and (b) how strong their impact is on the evolution of the global system. These questions can only be answered by combining theory, simulations, and spacecraft observations. A quantification of the different contributions of instabilities would be worthwhile, as recently provided by \citet{zhao22} for 
    Alfv\'en  waves in collisionless plasmas.
    \item  As we show through our quasi-linear model, resonant instabilities strongly depend on the exact shape of the distribution function near the parallel speed that fulfills Eq.~(\ref{rescond}) via $\hat Gf_{0j}$. While temperature anisotropies and reflectional asymmetries in the distribution function each represent non-equilibrium features, natural plasmas are likely to exhibit a combination of both. Some recent studies combine these drivers in their analyses of electron-driven instabilities \citep{lazar18,vasko20,shaaban20}. While most theoretical descriptions characterise the instability-driving non-equilibrium features with prescribed distributions (e.g., with drifting bi-Maxwellian or bi-$\kappa$ distributions), it is more reliable (and potentially necessary) to evaluate the stability of the actual distribution functions without relying on distribution models \citep{dum80}. Modern numerical tools such as LEOPARD \citep{astfalk17} and ALPS \citep{verscharen18} exist that are capable of this evaluation; however, a systematic application to measured electron distributions is still outstanding \citep[for notable exceptions, see][]{husidic20,page21,schroeder21}.
    \item  Parker Solar Probe will continue to measure electron distribution functions in close proximity to the Sun. Solar Orbiter will measure electron distribution functions outside the ecliptic plane. With its modern sets of instrumentation, both missions will generate unprecedented amounts of solar-wind electron data in combination with measurements of fluctuations in the magnetic and electric fields. These data will resolve some of the listed science challenges here but also pose new questions about the action and role of electron-driven instabilities in the solar wind. Moreover, the space-plasma community is planning future missions, like the mission proposal Debye, dedicated to electron-scale kinetics and its impact on the global plasma system \citep{wicks19,verscharen21}. 
\end{enumerate}

\section*{Conflict of Interest Statement}

The authors declare that the research was conducted in the absence of any commercial or financial relationships that could be construed as a potential conflict of interest.

\section*{Author Contributions}

D.V.~took the lead role of the coordination and writing of the article.
J.H., V.K.J., {\v{S}}.{\v{S}}., and P.L.W.~contributed to the writing of the paper, focusing on the comparison to observations.  E.B., M.E.I., and A.M.~contributed to the writing of the paper, focusing on the numerical simulations of electron-driven instabilities.
B.D.G.C., V.P., I.V., and M.V.~contributed to the writing of the paper, the broader context, and by completing the list of references.

\section*{Funding}
D.V.~is supported by the UK Science and Technology Facilities Council (STFC) Ernest Rutherford Fellowship ST/P003826/1, and STFC Consolidated Grants ST/S000240/1 and ST/W001004/1.
B.D.G.C.~acknowledges support from NASA grant 80NSSC19K0829 and from  the Parker Solar Probe  FIELDS Experiment through NASA grant NNN06AA01C.
J.H.~is supported by the Parker Solar Probe mission through the SWEAP contract NNN06AA01C. M.E.I.~acknowledges support from the German Science Foundation DFG within the Collaborative Research Center SFB1491.

\section*{Acknowledgments}
D.V.~is very grateful for many discussions with Peter Gary at various occasions about linear Vlasov--Maxwell theory, plasma instabilities, and turbulence in the solar wind. In particular, conversations with Peter about the oblique fast-magnetosonic/whistler instability at the 2014 SHINE Workshop in Telluride CO and more broadly about electron kinetics at the 2015 SHINE Workshop in Stowe VT sparked D.V.'s interest in electron-driven instabilities in the solar wind. D.V.~also appreciates discussions about the theory of the oblique fast-magnetosonic/whistler instability with Stuart Bale, Eliot Quataert, and Chadi Salem at the 2013 HTP team meeting in Berkeley CA.

The authors acknowledge insightful discussions within the International Team ``Heliospheric Energy Budget: From Kinetic Scales to Global Solar Wind Dynamics'' at the International Space Science Institute (ISSI) in Bern led by M.~E. Innocenti and A.~Tenerani. This work was discussed at the ``Joint Electron Project'' at MSSL.

\bibliographystyle{frontiersinSCNS_ENG_HUMS} 
\bibliography{frontiers}

\end{document}